\documentclass{article}
\usepackage[left=3.5cm,right=3.5cm,top=3.5cm,bottom=3.5cm,bindingoffset=0cm]{geometry}

\usepackage{longtable}
\usepackage{mathtools}
\usepackage{cases}

\usepackage{upgreek}
\usepackage{epsfig}
\usepackage{shadethm}
\usepackage[mathscr]{euscript}
\usepackage{amsxtra}
\usepackage{amssymb}
\usepackage{calligra}
\usepackage{calrsfs}

\usepackage[colorlinks=true,linkcolor=blue, citecolor=blue, urlcolor=blue]{hyperref}

\usepackage[firstinits=true, parentracker=true, doi=true, sorting=nyt,isbn=false, url=false, eprint=true, ]{biblatex}

\usepackage{chngcntr}
\counterwithout{figure}{section}

\renewcommand{\url}[1]{}

\usepackage{chngcntr}
\counterwithout{figure}{section}
\usepackage{float}
\usepackage{lipsum}
\usepackage[figuresright]{rotating}

\usepackage{mdframed}
\usepackage{lineno}

\newtheorem{definition}{Definition}[section]\newtheorem{remark}[definition]{Remark}

\numberwithin{equation}{section}
\DeclareMathAlphabet{\mathpzc}{OT1}{pzc}{m}{it}
\DeclareMathOperator{\sqh}{\mathpzc{p}}

\definecolor{shadecolor}{gray}{0.95}

\newcommand{\la}{\mathrm{b}}
\newcommand{\su}{\mathtt{u}}
\usepackage{amsmath,environ}
\NewEnviron{eqn}{\begin{equation}\begin{split}
 \BODY
\end{split}\end{equation}
}
\def\sbfrac#1#2{\ensuremath{\genfrac{}{}{0pt}1{\raisebox{-.5ex}{\ensuremath{#1}}}{\raisebox{.5ex}{\ensuremath{#2}}}}}
\newcommand{\beqans}{\begin{subequations}\begin{eqnarray}}
\newcommand{\eeqans}[1]{\end{eqnarray}\label{#1}\end{subequations}}
\newcommand{\beqan}{\begin{eqnarray}}
\newcommand{\eeqan}{\end{eqnarray}}
\DeclareMathOperator{\sq}{\mathpzc{q}}
\newcommand{\pd}[2]{\frac{\partial#1}{\partial#2}}
\newcommand*{\bfrac}[2]{\genfrac{}{}{0pt}{}{\raisebox{-.3em}{\scriptsize$#1$}}{\raisebox{.4em}{\scriptsize$#2$}}}

\begin{document}

\title{Transmission of waves across atomic step discontinuities in discrete nanoribbon structures}
\author{Basant Lal Sharma\thanks{Department of Mechanical Engineering, Indian Institute of Technology Kanpur, Kanpur, U. P. 208016, India ({bls@iitk.ac.in}).}}
\date{\today}
\maketitle
\begin{abstract}
{Scalar wave propagation across a semi-infinite step or step-like discontinuity on any one boundary of the square lattice waveguides is considered within nearest-neighbour interaction approximation. An application of the Wiener--Hopf method does yield an exact solution of the discrete scattering problem, using which, as the main result of the paper, the transmission coefficients for energy flux are obtained. 
It is assumed that a wave mode is incident from either side of the step and the question addressed is what fraction of incident energy is transmitted across the atomic step discontinuity. 
A total of ten configurations are presented that arise due to various placements of discrete Dirichlet and Neumann boundary conditions on the waveguide. 
Numerical illustrations of a measure of `conductance' are provided.}
\end{abstract}

\section*{Introduction}
\label{intro}
Historically, an understanding of the surface roughness induced scattering of {waves} has played a crucial role in the subject of mechanics, physics, and thermodynamics of solids \cite{Papadopoulos1957,Cahill1,Cahill2,Fellay,Ladik1987,Kosevich2008,Kosevichmulti,santamore2001effect,sanchez1998coexistence,sanchez1999reflection,Mujica1994_1,Virlouvet}.
Classically, research works on waveguide{s}, with scattering due to steps, have been carried out by numerous investigators due to their importance in elastic, acoustic, and electromagnetic wave propagation theory; some of the results are also obtained through the Wiener--Hopf technique \cite{Wiener,Noble,mittra1971analytical}.
On the other hand, the {discrete} models appear in many applications involving crystalline matter
with or without impurities \cite{kosevich}, thin films or monolayers \cite{Ohring2002711,
Gong,Ohtake}, etc. 
Within this context of surface induced phenomena, the present paper extends the applications of 
{recent framework developed for} discrete scattering problems 
\cite{Bls0, Bls1,Bls8pair1}. 
In contrast to the recent investigation of the wave transmission of surface localized anti-plane shear wave \cite{Victor_Bls_surf2} in lattice structures, motivated by certain discrete counterpart \cite{Victor_Bls_surf1} of Gurtin-Murdoch model \cite{GurtinMurdoch1975a,gurtin1978surface,steigmann1997plane,SteigmannOgden1999}, the present paper applies similar methodology to reveal the transmission characteristic associated with a close-up of discrete form of surface roughness.

As observed in this paper, the time harmonic response of lattice waveguides, modulo the presence of a step or a step-like discontinuity, 
is also remarkable. 
It turns out that there exist frequency intervals where certain measure of conductance \cite{Rego1998,schwab2000measurement}) is large and in some other intervals there is dominant effect of step induced scattering, depending on the boundary conditions as well. It can be envisaged that the techniques and results of the paper assume qualitative relevance to 
more complex {physical} systems \cite{marcuvitz1951waveguide,collin1991field,linton2001handbook, BurnsMilton,
kokubo2011waveguide}.
The lattice waveguides {also demonstrate a} quasi-one dimensional character, 
providing a pathway to venture into the effect of second dimension in physical space. 
The paper maintains focus on the effects due to presence of confinement, incorporated by the boundary conditions.
The derived results which can be obtained, more or less directly, 
based on the manipulations and expressions 
of the exact solution 
on infinite square lattice \cite{Bls0, Bls1} and lattice waveguides \cite{Bls9s}, 
have been omitted in the main paper. 
Throughout the paper, a useful role is played by Chebyshev polynomials \cite{MasonHand}.

\begin{figure}[ht]\centering{\includegraphics[width=\textwidth]{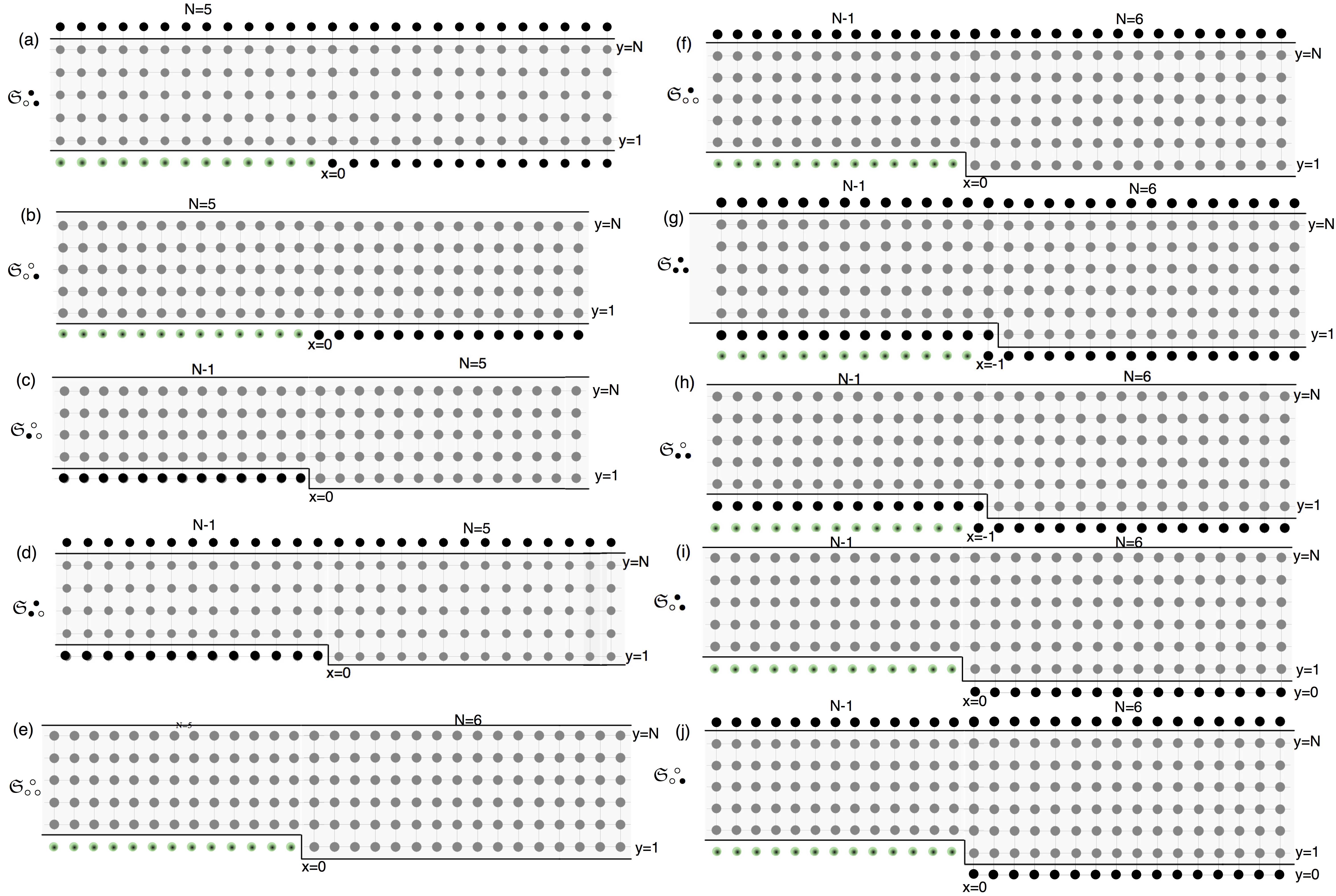}}
\caption{Waveguides of square lattice structure with step-like (a--d) discontinuity induced by a union of semi-infinite `discrete Neumann' and `discrete Dirichlet' (a, b) boundary and a semi-infinite `discrete Dirichlet' and `discrete Neumann' (c, d) boundary or a step (e--h) discontinuity on a `discrete Neumann' (e, f) and a `discrete Dirichlet' (g, h) boundary. }
\label{stepboundarywaveguides_sq}\end{figure}

\section{Square lattice waveguide}
Consider a strip with 
${\mathtt{N}}$ number of rows of the square lattice structure, $\{\nu^i\hat{\mathbf{{e}}}_{i}\big| \nu^1\in{\mathbb{Z}}, \nu^2\in\mathbb{Z}_1^{{\mathtt{N}}}\}.$
Each particle of such discrete nanoribbon structure is assumed to possess unit mass, and, interaction with atmost its four nearest neighbours through linearly elastic identical (massless) bonds with a spring constant $1/\la^2$. The notation follows that introduced for an infinite lattice \cite{Bls0, Bls1} and the square lattice waveguides \cite{Bls9,Bls9s}.
Due to nature of the problem, the displacement of a particle, located at the site indexed by its {lattice coordinates} $({\mathtt{x}}, {\mathtt{y}})\in{{\mathbb{Z}^2}}$ in ${\mathfrak{S}}_{{\mathtt{N}}}$ and denoted by ${\su}_{{\mathtt{x}}, {\mathtt{y}}}$, is complex valued. The equation of motion of the particles in ${\mathfrak{S}}_{{\mathtt{N}}}$
away from the upper and lower boundaries is
\begin{eqn}
\la^2\ddot{\su}_{{\mathtt{x}}, {\mathtt{y}}}&={\triangle}{\su}_{{\mathtt{x}}, {\mathtt{y}}}, \text{where }{\triangle}{\su}_{{\mathtt{x}}, {\mathtt{y}}}{:=}{\su}_{{\mathtt{x}}+1, {\mathtt{y}}}+{\su}_{{\mathtt{x}}-1, {\mathtt{y}}}+{\su}_{{\mathtt{x}}, {\mathtt{y}}+1}+{\su}_{{\mathtt{x}}, {\mathtt{y}}-1}-4{\su}_{{\mathtt{x}}, {\mathtt{y}}}. 
\label{newtoneq_sq}
\end{eqn}
The equation of motion at upper (corresponding to ``$-$'') and (each left or right part of the) lower (corresponding to ``$+$'') boundary rows studied in this paper is \text{either }
\begin{subequations}
\begingroup
\addtolength{\jot}{0em}
\begin{align}
\la^2\ddot{\su}_{{{\mathtt{x}}}, {{\mathtt{y}}}}&={\su}_{{{\mathtt{x}}}+1, {{\mathtt{y}}}}+{\su}_{{{\mathtt{x}}}-1, {{\mathtt{y}}}}+{\su}_{{{\mathtt{x}}}, {{\mathtt{y}}}\mp1}-4{\su}_{{{\mathtt{x}}}, {{\mathtt{y}}}}
&\quad&\text{(${\bullet}$)}\label{bc_sq_X}\\ \text{or }\la^2\ddot{\su}_{{{\mathtt{x}}}, {{\mathtt{y}}}}&={\su}_{{{\mathtt{x}}}+1, {{\mathtt{y}}}}+{\su}_{{{\mathtt{x}}}-1, {{\mathtt{y}}}}+{\su}_{{{\mathtt{x}}}, {{\mathtt{y}}}\mp1}-3{\su}_{{{\mathtt{x}}}, {{\mathtt{y}}}}&\quad&\text{(${\circ}$)}.\label{bc_sq_R} 
\end{align}
\endgroup
\end{subequations}
In the context of the waveguides with step (or step-like) discontinuity studied in this paper,
in particular, on the upper free boundary (resp. on row adjacent to fixed boundary) \eqref{bc_sq_R} holds (resp. \eqref{bc_sq_X}) with $-$ sign; 
on the lower free boundary (resp. on row adjacent to fixed boundary) \eqref{bc_sq_R} holds (resp. \eqref{bc_sq_X}) for ${\mathtt{x}}<0$ and/or ${\mathtt{x}}\ge0$ 
with $+$ sign. 
The same geometric structure can be also interpreted as a junction of two semi-infinite waveguides of either of the four kinds as stated in \cite{Bls9} (briefly recalled in Appendix \ref{app_recallbifwavesq_Chebdef}). The ten configurations shown in Fig. \ref{stepboundarywaveguides_sq} as a graphical schematic illustration of the boundary conditions. 
Again the notation is self-explanatory as the superscript on ${\mathfrak{S}}$ represents the upper boundary condition and the subscript represents the pair of two semi-infinite lower boundary conditions.
Fig. \ref{stepboundarywaveguides_sq} illustrates schematically the step discontinuity induced by a union of semi-infinite `discrete Neumann' and `discrete Dirichlet' (i, j) boundary.
The green dots are unconnected hypothetical sites used in mathematical formulation. 
The sites contained in the shaded region bounded by the solid lines belong to the waveguide, wherein the sites adjacent to the solid lines belong to ${\Sigma}$ (such that \eqref{bc_sq_X} or \eqref{bc_sq_R} holds).

Suppose ${\su}^{{\text{in}}}$ describes the { incident wave mode} (indexed by natural number ${{\kappa}}^{{\text{in}}}$ \cite{Bls9}) with frequency $\omega$ and a { lattice wave number} ${\upkappa}_x$ along ${\mathtt{x}}$ coordinate. The ordinary frequency $\omega$ (aside from a factor due to scaling of time \cite{Bls0}) corresponds to a propagating mode lying in the pass band of lattice waveguide { either on the left or right side} of the step discontinuity. It is assumed that ${\su}^{{\text{in}}}$ is given by an expression of the form
\begin{eqn}
{\su}_{{\mathtt{x}}, {\mathtt{y}}}^{{\text{in}}}{:=}{{\mathrm{A}}}{{a}}_{({{\kappa}}^{{\text{in}}}){{\mathtt{y}}}}e^{-i{\upkappa}_x {\mathtt{x}}-i\omega t}, \forall ({\mathtt{x}}, {\mathtt{y}})\in{{\mathbb{Z}^2}N}, 
\label{uinc_sq}
\end{eqn}
where ${{\mathrm{A}}}\in\mathbb{C}$ is constant. The symbol ${{a}}_{({{\kappa}}^{{\text{in}}})\cdot}$ represents the mode shape corresponding to the specific wave mode indexed by ${{\kappa}}^{{\text{in}}}$ (see \cite{Bls9} or Appendix C of \cite{Bls9s} for the detailed expressions). 
In the remaining text, the explicit time dependence factor, $e^{-i\omega t}$, is suppressed. 

Let the { lattice frequency} ${\upomega}$ be defined as ${\upomega}{:=} \la\omega,$
so that ${\upomega}^2=4(\sin^2{\frac{1}{2}}{\upkappa}_x+\sin^2{\frac{1}{2}}{\upeta_{\kappa^{\text{in}}}}), {\upkappa}_x\in [-\pi, \pi],$
where ${\upeta_{\kappa^{\text{in}}}}$ is interpreted as the vertical component of an incident wave vector in (bulk) lattice. Clearly, ${\upeta_{\kappa^{\text{in}}}}$ depends on the confinement effect of the boundary conditions \cite{Bls9}. The total displacement ${\su}^{{t}}$ of an arbitrary particle in each lattice waveguide is a sum of the incident wave displacement ${\su}^{{\text{in}}}$ and the scattered wave displacement ${\su}^{{s}}$. 
Henceforth, the letter ${\su}$ is used in place of ${\su}^{{s}}$. The total displacement field ${\su}^{{t}}$ ($={\su}_{{\mathtt{x}}, {\mathtt{y}}}^{{\text{in}}}+{\su}_{{\mathtt{x}}, {\mathtt{y}}}$) satisfies the discrete Helmholtz equation
\begin{eqn}
{\triangle}{\su}^{{t}}_{{\mathtt{x}}, {\mathtt{y}}}+{\upomega}^2{\su}^{{t}}_{{\mathtt{x}}, {\mathtt{y}}}&=0, ({\mathtt{x}}, {\mathtt{y}})\in{{\mathbb{Z}^2}N}\setminus{{\Sigma}} 
\text{ with }
{\upomega}={\upomega}_1+i{\upomega}_2, {\upomega}_2>0,
\label{dHelmholtz_sq}
\end{eqn}
where ${{\Sigma}}$ is boundary of the waveguide (as enumerated in Fig. \ref{stepboundarywaveguides_sq}). 
The discrete Fourier transform \cite{jury,Slepyanbook} ${\su}_{{\mathtt{y}}}^F: \mathbb{C}\to\mathbb{C}$ of $\{{\su}_{{\mathtt{x}}, {\mathtt{y}}}\}_{{\mathtt{x}}\in\mathbb{Z}}$ (along the ${\mathtt{x}}$ axis) is defined by
\begin{eqn}
{\su}_{{\mathtt{y}}}^F&={\su}_{{\mathtt{y}};+}+{\su}_{{\mathtt{y}};-}, \text{ where }
{\su}_{{\mathtt{y}};+}({{z}})=\sum\nolimits_{{\mathtt{x}}=0}^{+\infty} {\su}_{{\mathtt{x}}, {\mathtt{y}}}{{z}}^{-{\mathtt{x}}}, 
{\su}_{{\mathtt{y}};-}({{z}})=\sum\nolimits_{{\mathtt{x}}=-\infty}^{-1} {\su}_{{\mathtt{x}}, {\mathtt{y}}}{{z}}^{-{\mathtt{x}}}.\label{unpm}\end{eqn}
The discrete Fourier transform ${\su}_{{\mathtt{y}}}^F$ of  $\{{\su}_{{\mathtt{x}}, {\mathtt{y}}}\}_{{\mathtt{x}}\in\mathbb{Z}}$ is well defined for all relevant values of (fixed) ${\mathtt{y}}$ \cite{Bls9s} in an annulus ${\mathscr{A}}$ in the complex plane (same as that defined in \cite{Bls9s}, ${{\mathit{R}}}_\pm$ are given in Appendix A of \cite{Bls9s}, namely, equation (A.1) ${{\mathit{R}}}_+=e^{-{\upkappa}_2}, {{\mathit{R}}}_-=e^{+{\upkappa}_2}$).

\section{Step or step-like discontinuity induced by a semi-infinite fixed or free boundary}
In this section the ten wave-guides with boundary containing a step (or step-like discontinuity), as shown schematically in Fig. \ref{stepboundarywaveguides_sq}, are considered one by one. It is assumed that the wave is incident from the portion ahead of the step (the other case of incidence is treated afterwards). 

\subsection{Incidence from the broad portion of the step discontinuity}
\subsubsection{Case: (a)}
\label{crack_sq_X}
In this case the particles at ${\mathtt{x}}\in\mathbb{Z}, {\mathtt{y}}={{\mathtt{N}}}+1$ and ${\mathtt{x}}\ge0, {\mathtt{y}}=0$ {are} fixed while those at ${\mathtt{x}}<0, {\mathtt{y}}=1$ {belong} to a free boundary, as shown in Fig. \ref{stepboundarywaveguides_sq}(a). Thus, ${\mathfrak{S}\hspace{-.4ex}}{\mathbin{\substack{{\bullet}\\{\circ}{\bullet}}}}$ is the waveguide constructed by a junction between ${\mathfrak{S}{\sbfrac{\bullet}{\circ}}}$ and ${\mathfrak{S}{\sbfrac{\bullet}{\bullet}}}$ of width ${\mathtt{N}}$ (recall the notation described in Appendix \ref{app_recallbifwavesq_Chebdef} or \cite{Bls9}). 
{Using the general solution \eqref{dHelmholtz_sq} in the square lattice \eqref{gensol_sq} and upper boundary condition \eqref{bc_sq_X},}
\beqans
{\su}_{{\mathtt{y}}}^F ={\su}^F_1{\Lambda}_{{\mathtt{y}}}({\mathtt{N}},{\bullet};1), {{\mathtt{y}}}\in\mathbb{Z}_1^{{\mathtt{N}}}, \label{bulk_k_sq_X}\\
\text{where }
{\Lambda}_{{\mathtt{y}}}(n,{\bullet};1)=\frac{{{\lambda}}^{{\mathtt{y}}-1}-{{\lambda}}^{2{n}-{{\mathtt{y}}}+1}}{1-{{\lambda}}^{2{n}}}, \label{U_k_sq_X}
\eeqans{ubulk_k_sq_X}
where ${\lambda}$ \cite{Slepyanbook,Bls0} is defined by \eqref{lambdadef_sq} and the variable of these complex functions (${z}$) lies in ${\mathscr{A}}$.
According to \eqref{bulk_k_sq_X}, it is easy to notice that ${\su}_{\cdot, {\mathtt{y}}}$ equals ${\su}_{\cdot, 1}$ for ${\mathtt{y}}=1$, and is identically zero for ${\mathtt{y}}={{\mathtt{N}}}+1$ (due to the presence of {normal mode factor} in \eqref{uinc_sq}, ${{\su}^{{\text{in}}}_{\cdot, 0}}=0,$ ${{\su}^{{\text{in}}}_{\cdot, {\mathtt{N}}+1}}=0$). 
As an analogue of \eqref{dHelmholtz_sq} for ${\mathtt{y}}=1,$
\begin{eqn}
&(-{\upomega}^2{\su}_{{{\mathtt{x}}}, 1}-{{{b}^2}}{{{\mathit{v}}}^{{t}}}_{{\mathtt{x}}})=-({\su}_{{\mathtt{x}}, 0}^{{\text{in}}}-{\su}_{{\mathtt{x}}, 1}^{{\text{in}}})
+({\su}_{{{\mathtt{x}}}+1, 1}+{\su}_{{{\mathtt{x}}}-1, 1}+{\su}_{{{\mathtt{x}}}, 2}-3{\su}_{{{\mathtt{x}}}, 1}),
\label{dimnewtoneq_k_sq_X}
\end{eqn}
where ${{{\mathit{v}}}^{{t}}}_{{\mathtt{x}}}$ is the total force that acts on the particle $({{\mathtt{x}}}, 1)$ (from below).
As a consequence of the step-like discontinuity between a semi-infinite discrete Neumann and discrete Dirichlet boundary, ${{{\mathit{v}}}^{{t}}}_{{{\mathtt{x}}}}{:=}0, {{\mathtt{x}}}<0, $ so that ${{\mathit{v}}}^{{t}}_{-}=0$, according to the minus function defined in \eqref{unpm} (analytic on ${\mathscr{A}}$ and in its interior). 
By an application of the Fourier transform \eqref{unpm}, 
it follows that
\beqans
{{{b}^2}}{{{\mathit{v}}}^{{t}}}^F&=&-{{{\mathtt{v}}}^{{\text{in}}}}^F+(({{\mathpzc{Q}}}-1) {\su}^F_1-{\su}^F_2),
\label{uF_k_sq_X_veq}\\
\text{where }
{{{b}^2}}{{{\mathit{v}}}^{{t}}}^F&=&-({{\su}_{{\mathtt{x}}, 1}^{{t}}}-{{\su}_{{\mathtt{x}}, 0}^{{t}}}){\mathit{H}}({\mathtt{x}}), 
\\{{{\mathtt{v}}}^{{\text{in}}}_{{\mathtt{x}}}}&{:=}&{{\su}_{{\mathtt{x}}, 1}^{{\text{in}}}}-{{\su}_{{\mathtt{x}}, 0}^{{\text{in}}}}={{\su}_{{\mathtt{x}}, 1}^{{\text{in}}}}, {\mathtt{x}}\in\mathbb{Z},
\label{vinc_def_X}
\eeqans{auxeqcase_a}
and ${\mathpzc{Q}}$ is defined by \eqref{dHelmholtzF_sq} (see also \cite{Bls0,Bls9s}, and a plethora of works cited in \cite{Slepyanbook} using similar notation and definitions).
Simplifying and rewriting \eqref{uF_k_sq_X_veq}, it is found that
\begin{eqn}
(({{\mathpzc{Q}}}-1) {\su}^F_1-{\su}^F_2)+{\su}_{1;+}={\su}^{{\text{in}}}_{1;-},
\label{uF_k_sq_X}
\end{eqn}
where ${\su}_{1;+}$ is defined according to the plus function defined in \eqref{unpm} (analytic on ${\mathscr{A}}$ and in its exterior). 
Using the definition of Chebyshev polynomial of the second kind \cite{Chebyshev00,MasonHand}, the following identity holds 
\begin{eqn}
{\lambda}^{-n}-{\lambda}^{n}=({\lambda}^{-1}-{\lambda}){\mathtt{U}}_{n-1}({\frac{1}{2}}({\lambda}+{\lambda}^{-1})), 0<n\in\mathbb{Z},
\label{identity1eq}
\end{eqn}
on ${\mathscr{A}}.$
The expression of ${\su}^F_{2}$ in terms of ${\su}^F_{1}$, from \eqref{bulk_k_sq_X} and \eqref{identity1eq}, is found to be
\begin{eqn}
{\su}^F_{2}={\su}^F_{1}{\mathpzc{V}}_{{\mathtt{N}}}, \quad{\mathpzc{V}}_{{\mathtt{N}}}=\frac{{\mathtt{U}}_{{\mathtt{N}}-2}({\vartheta})}{{\mathtt{U}}_{{\mathtt{N}}-1}({\vartheta})},
\label{u2Fu1F_sq_X}
\end{eqn}
where (and also throughout the paper) the argument of Chebyshev polynomials ${\vartheta}$ is defined by
\begin{eqn}
{\vartheta}{:=}{\vartheta}({z})=\tfrac{1}{2}{\mathpzc{Q}}({{z}}), {{z}}\in\mathbb{C}.
\label{chebx_sq}\end{eqn}
The equation \eqref{chebx_sq} utilizes the definition of ${\mathpzc{Q}}$ provided as \eqref{q2_sq} and the related fact that ${\mathpzc{Q}}={\lambda}+{\lambda}^{-1}$, where ${{\lambda}}$ is given by \eqref{lambdadef_sq}. 
After the substitution of \eqref{u2Fu1F_sq_X} in \eqref{uF_k_sq_X}, a scalar Wiener--Hopf equation \cite{Noble} can be constructed for ${\su}_1^F={\su}_{1;+}+{\su}_{1;-}$ (the only unknown in \eqref{bulk_k_sq_X}) as 
\begin{eqn}
{\mathpzc{L}}{\su}_{1;+}+{\su}_{1;-}&=-(1-{\mathpzc{L}})\su^{{\text{in}}}_{1;-},
\label{WHKeq_sq_X}
\end{eqn}
on ${\mathscr{A}},$
where the function ${\mathpzc{L}}$ is the Wiener--Hopf kernel given by 
\begin{eqn}
{{\mathpzc{L}}}{:=}(1-\frac{1}{{\mathpzc{Q}}-{\mathpzc{V}}_{{\mathtt{N}}}})^{-1}.
\label{LNgen_sq_X}
\end{eqn}
As a consequence of 
the identities involving Chebyshev polynomials \cite{Chebyshev00,MasonHand} (see also Appendix D of \cite{Bls9s} for the latter, see also \cite{Bls9}), \eqref{LNgen_sq_X} can be simplified to the form
\begin{eqn}
{{\mathpzc{L}}}=\frac{{\mathtt{U}}_{{\mathtt{N}}}({\vartheta})}{{\mathtt{U}}_{{\mathtt{N}}}({\vartheta})-{\mathtt{U}}_{{\mathtt{N}}-1}({\vartheta})}=\frac{{\mathtt{U}}_{{\mathtt{N}}}({\vartheta})}{{\mathtt{V}}_{{\mathtt{N}}}({\vartheta})}=\frac{{\mathscr{N}}}{{\mathscr{D}}}.
\label{Lk_sq_X}
\end{eqn}

\subsubsection{Case: (b)}
\label{crack_sq_RX}
{In this case} the particles at ${\mathtt{x}}\in\mathbb{Z}, {\mathtt{y}}={{\mathtt{N}}}$ {belong} to a free boundary and ${\mathtt{x}}\ge0, {\mathtt{y}}=1$ {are} fixed while those at ${\mathtt{x}}<0, {\mathtt{y}}=1$ also {belong} to a free boundary, as shown in Fig. \ref{stepboundarywaveguides_sq}(b). Thus, ${\mathfrak{S}\hspace{-.4ex}}{\mathbin{\substack{\circ\\\circ\bullet}}}$ is the waveguide constructed by a junction between ${\mathfrak{S}{\sbfrac{\circ}{\circ}}}$ and ${\mathfrak{S}{\sbfrac{\circ}{\bullet}}}$ of width ${\mathtt{N}}$ (recall the notation described in Appendix \ref{app_recallbifwavesq_Chebdef} or \cite{Bls9}). 
For this choice of upper boundary \eqref{bc_sq_R}, similar to \eqref{U_k_sq_X}, let
\begin{subequations}
\begin{eqn}
{\Lambda}_{{\mathtt{y}}}(n,{\circ};1)=\frac{{{\lambda}}^{{\mathtt{y}}}+{{\lambda}}^{2{n}-{{\mathtt{y}}}+1}}{{{\lambda}}+{{\lambda}}^{2{n}}},
\label{U_sq_RX}
\end{eqn}
The ansatz for solution can be easily found to be 
\begin{eqn}
{\su}_{{\mathtt{y}}}^F ={\su}^F_1{\Lambda}_{{\mathtt{y}}}({\mathtt{N}},{\circ};1), \label{bulk_k_sq_RX}
\end{eqn}
\end{subequations}
on ${\mathscr{A}},$
with ${{\lambda}}$ given by \eqref{lambdadef_sq}. Analogous to \eqref{u2Fu1F_sq_X}, the relation between ${\su}_{2}^F$ and ${\su}^F_1$ is  as
\begin{eqn}
{\su}_{2}^F ={\su}^F_1{{\mathpzc{V}}_{{\mathtt{N}}}}, \quad{{\mathpzc{V}}_{{\mathtt{N}}}}=\frac{{\mathtt{V}}_{{\mathtt{N}}-2}({\vartheta})}{{\mathtt{V}}_{{\mathtt{N}}-1}({\vartheta})}.
\label{u2Fu1F_sq_RX}
\end{eqn}
Further, using the manipulations described in \S\ref{crack_sq_X}, \eqref{WHKeq_sq_X} is obtained (for the only unknown ${\su}^F_1$ in \eqref{bulk_k_sq_RX}) with the Wiener--Hopf kernel ${{\mathpzc{L}}}$ \eqref{LNgen_sq_X} as
\begin{eqn}
{{\mathpzc{L}}}&=\frac{{\mathtt{U}}_{{\mathtt{N}}}({\vartheta})-{\mathtt{U}}_{{\mathtt{N}}-1}({\vartheta})}{{\mathtt{U}}_{{\mathtt{N}}}({\vartheta})-{\mathtt{U}}_{{\mathtt{N}}-1}({\vartheta})-({\mathtt{U}}_{{\mathtt{N}}-1}({\vartheta})-{\mathtt{U}}_{{\mathtt{N}}-2}({\vartheta}))}=\frac{{\mathtt{V}}_{{\mathtt{N}}}({\vartheta})}{{\mathpzc{H}}{\mathtt{U}}_{{\mathtt{N}}-1}({\vartheta})}=\frac{{\mathscr{N}}}{{\mathscr{D}}},
\label{Lk_sq_RX}
\end{eqn}
on ${\mathscr{A}}.$
Recall that ${\mathpzc{H}}$ ($={\mathpzc{Q}}-2$) is defined by \eqref{HR_sq}.

\subsubsection{Case: (c)}
\label{slit_sq_R}
{In this case}, particles at ${\mathtt{x}}\in\mathbb{Z}, {\mathtt{y}}={{\mathtt{N}}}$, ${\mathtt{x}}\ge0, {\mathtt{y}}=1$ {belong} to a free boundary while those at ${\mathtt{x}}<0, {\mathtt{y}}=1$ are fixed, as shown in Fig. \ref{stepboundarywaveguides_sq}(c). Thus, ${\mathfrak{S}\hspace{-.4ex}}{\mathbin{\substack{\circ\\\bullet\circ}}}$ is waveguide constructed by junction between ${\mathfrak{S}{\sbfrac{\circ}{\bullet}}}$ and ${\mathfrak{S}{\sbfrac{\circ}{\circ}}}$ of widths ${\mathtt{N}}-1$ and ${\mathtt{N}}$, respectively.
For this choice of upper boundary \eqref{bc_sq_R}, using \eqref{U_sq_RX}, ansatz for solution can be easily found to be \eqref{bulk_k_sq_RX}.
In place of \eqref{dHelmholtz_sq} for ${\mathtt{y}}=1$ is
\begin{subequations}
\begin{eqn}
-{\upomega}^2{\su}_{{{\mathtt{x}}}, 1}&=({\su}_{{{\mathtt{x}}}-1, 1}+{\su}_{{{\mathtt{x}}}+1, 1}+{\su}_{{{\mathtt{x}}}, 2}-3{\su}_{{{\mathtt{x}}}, 1}){\mathit{H}}({\mathtt{x}})
+{\upomega}^2{\su}^{{\text{in}}}_{{{\mathtt{x}}}, 1}{\mathit{H}}(-1-{\mathtt{x}}).
\label{dimnewtoneqslit_sq_R}
\end{eqn}
By an application of the Fourier transform \eqref{unpm}, 
it follows that
\begin{eqn}
-{\upomega}^2{\su}_{1;-}({z})&=({z}+{z}^{-1}-3+{\upomega}^2){\su}_{1;+}({z})
+{\su}_{2;+}({z})+{\upomega}^2{\su}^{{\text{in}}}_{1;-}({z})+{{\mathpzc{W}}}({z}),\label{uF_c_sq_R}
\end{eqn}
\label{auxeqncase_c}
\end{subequations}
for ${z}\in{\mathscr{A}}.$
In \eqref{uF_c_sq_R}, using 
${\mathpzc{Q}}$ \eqref{dHelmholtzF_sq}, it is found that (on ${\mathscr{A}}$)
\beqans
{{\mathpzc{W}}}+({{\mathpzc{Q}}}-1){\su}_{1;-}&=&({{\mathpzc{Q}}}-1){\su}^F_1-{\su}_{2{;}+}, 
\label{sliteq_sq_R}\\
{\su}_{1; -}&=&-\su^{{\text{in}}}_{1;-}, 
\label{u1n_R}\\
\text{where }{{\mathpzc{W}}}({{z}})&{:=}&{\su}_{-1, 1}-{{z}} {\su}_{0, 1}, {{z}}\in\mathbb{C}\label{q2beta_R},
\eeqans{u0_R}
(similar format of equations arise in \cite{Bls1}). 
After the substitution of \eqref{bulk_k_sq_RX}, to obtain expression of ${\su}^F_{1}$ in terms of ${\su}^F_{2}$, i.e. \eqref{u2Fu1F_sq_RX},
after {a rearrangement of} \eqref{sliteq_sq_R}, and the transform \eqref{unpm}, it follows that ${\su}^F_{2}$ (as the only unknown in \eqref{bulk_k_sq_RX} after substitution of \eqref{u2Fu1F_sq_RX}) satisfies 
\beqans
{{\mathpzc{L}}}{\su}_{2;+}+{\su}_{2;-}&=&(1-{{\mathpzc{L}}})({{\mathpzc{W}}}-({{\mathpzc{Q}}}-1){\su}^{{\text{in}}}_{1;-}) \text{ on }{\mathscr{A}},\label{WHCeq_sq_R}\\
\text{where }
{{\mathpzc{L}}}&=&1-\frac{{\mathpzc{V}}_{{\mathtt{N}}}}{{\mathpzc{Q}}-1}. \label{WHkernel_sq_R}
\eeqans{WHeqslitfull_sq_R}
In the Wiener--Hopf equation \eqref{WHCeq_sq_R}, further, the kernel ${{\mathpzc{L}}}$ can be expressed as
\begin{eqn}
{{\mathpzc{L}}}&=\frac{{\mathpzc{H}}}{{\mathpzc{Q}}-1}\frac{{\mathtt{U}}_{{\mathtt{N}}-1}}{{\mathtt{V}}_{{\mathtt{N}}-1}}=\frac{{\mathscr{N}}}{{\mathscr{D}}}=\frac{{\mathscr{N}}}{({\mathpzc{Q}}-1)\mathring{{\mathscr{D}}}}.
\label{Lc_sq_R}
\end{eqn}

\subsubsection{Case: (d)}
\label{slit_sq_XR}
{In this case} the particles at ${\mathtt{x}}\in\mathbb{Z}, {\mathtt{y}}={{\mathtt{N}}}$ as well as at ${\mathtt{x}}<0, {\mathtt{y}}=1$ are fixed while those at ${\mathtt{x}}\ge0, {\mathtt{y}}=1$ {belong} to a free boundary, as shown in Fig. \ref{stepboundarywaveguides_sq}(d). Thus, ${\mathfrak{S}\hspace{-.4ex}}{\mathbin{\substack{\bullet\\\bullet\circ}}}$ is the waveguide constructed by a junction between ${\mathfrak{S}{\sbfrac{\bullet}{\bullet}}}$ and ${\mathfrak{S}{\sbfrac{\bullet}{\circ}}}$ of widths ${\mathtt{N}}-1$ and ${\mathtt{N}}$, respectively.
For this choice of upper boundary \eqref{bc_sq_X}, the ansatz for solution can be easily found to be \eqref{U_k_sq_X}. In particular, following \S\ref{slit_sq_R}, the expression of ${\su}^F_{1}$ in terms of ${\su}^F_{2}$ is \eqref{u2Fu1F_sq_X}, 
which leads to \eqref{sliteq_sq_R}, \eqref{u1n_R}, and \eqref{q2beta_R}, and eventually the Wiener--Hopf equation \eqref{WHCeq_sq_R} (for the only unknown in \eqref{U_k_sq_X} after substitution of \eqref{u2Fu1F_sq_X}), where ${{\mathpzc{L}}}$ \eqref{WHkernel_sq_R} can be simplified to obtain
\begin{eqn}
{{\mathpzc{L}}}&=\frac{1}{{\mathpzc{Q}}-1}\frac{{\mathtt{V}}_{{\mathtt{N}}}}{{\mathtt{U}}_{{\mathtt{N}}-1}}=\frac{{\mathscr{N}}}{{\mathscr{D}}}=\frac{{\mathscr{N}}}{({\mathpzc{Q}}-1)\mathring{{\mathscr{D}}}}.
\label{Lc_sq_XR}
\end{eqn}

\subsubsection{Case: (e)}
\label{crack_sq_R}
{In this case} the particles at ${\mathtt{x}}\in\mathbb{Z}, {\mathtt{y}}={{\mathtt{N}}}$, ${\mathtt{x}}\ge0, {\mathtt{y}}=1$ as well as at ${\mathtt{x}}<0, {\mathtt{y}}=2$ {belong} to a free boundary, as shown in Fig. \ref{stepboundarywaveguides_sq}(e). Thus, ${\mathfrak{S}\hspace{-.4ex}}{\mathbin{\substack{\circ\\\circ\circ}}}$ is the waveguide constructed by a junction between ${\mathfrak{S}{\sbfrac{\circ}{\circ}}}$ and ${\mathfrak{S}{\sbfrac{\circ}{\circ}}}$ of widths ${\mathtt{N}}-1$ and ${\mathtt{N}}$, respectively.
For this choice of upper boundary \eqref{bc_sq_R}, using \eqref{U_sq_RX},
the ansatz for solution can be easily found to be \begin{eqn}{\su}_{{\mathtt{y}}}^F ={\su}^F_2{\Lambda}_{{\mathtt{y}}}({\mathtt{N}}-1,{\circ};1), {\mathtt{y}}\in\mathbb{Z}_2^{{\mathtt{N}}}.\label{bulk_k_sq_R}\end{eqn}
Instead of \eqref{dHelmholtz_sq} for ${\mathtt{y}}=2, 1,$
\beqans
(-{\upomega}^2{\su}_{{{\mathtt{x}}}, 2}-{{{b}^2}}{{{\mathit{v}}}^{{t}}}_{{\mathtt{x}}})=-({\su}_{{\mathtt{x}}, 1}^{{\text{in}}}-{\su}_{{\mathtt{x}}, 2}^{{\text{in}}})
+({\su}_{{{\mathtt{x}}}+1, 2}+{\su}_{{{\mathtt{x}}}-1, 2}+{\su}_{{{\mathtt{x}}}, 3}-3{\su}_{{{\mathtt{x}}}, 2}), {\mathtt{x}}\in\mathbb{Z},\label{dimnewtoneqcracku_sq_R}\\
(-{\upomega}^2{\su}_{{{\mathtt{x}}}, 1}+{{{b}^2}}{{{\mathit{v}}}^{{t}}}_{{\mathtt{x}}})=({\su}_{{\mathtt{x}}, 1}^{{\text{in}}}-{\su}_{{\mathtt{x}}, 2}^{{\text{in}}})
+({\su}_{{{\mathtt{x}}}-1, 1}-{\su}_{{{\mathtt{x}}}, 1}){\mathit{H}}({\mathtt{x}}-1)+({\su}_{{{\mathtt{x}}}+1, 1}-{\su}_{{{\mathtt{x}}}, 1}){\mathit{H}}({\mathtt{x}})\notag\\
-({\su}^{{\text{in}}}_{{{\mathtt{x}}}-1, 1}-{\su}^{{\text{in}}}_{{{\mathtt{x}}}, 1}){\updelta}_{{\mathtt{x}},0}, {\mathtt{x}}\ge0,\label{dimnewtoneqcrackl_sq_R}
\eeqans{dimnewtoneqcrack_sq_R}
where ${{{\mathit{v}}}^{{t}}}_{{\mathtt{x}}}$ is the total force that acts on the particle $({{\mathtt{x}}}, 2)$, 
i.e.
\begin{eqn}
{{{b}^2}}{{{\mathit{v}}}^{{t}}}^F=-({{\su}_{{\mathtt{x}}, 2}^{{t}}}-{{\su}_{{\mathtt{x}}, 1}^{{t}}}){\mathit{H}}({\mathtt{x}}), \quad {{{\mathtt{v}}}^{{\text{in}}}_{{\mathtt{x}}}}{:=}{{\su}_{{\mathtt{x}}, 2}^{{\text{in}}}}-{{\su}_{{\mathtt{x}}, 1}^{{\text{in}}}}, {\mathtt{x}}\in\mathbb{Z},
\label{vinc_def_R}
\end{eqn}
(similar to \eqref{vinc_def_X}). By an application of the Fourier transform \eqref{unpm}, 
it follows that
\begin{eqn}
-{{\mathtt{v}}}_+-{{\mathtt{v}}}^{{\text{in}}}_+={{{b}^2}}{{{\mathit{v}}}^{{t}}}^F
&=-{{{\mathtt{v}}}^{{\text{in}}}}^F+(({{\mathpzc{Q}}}-1) {\su}^F_2-{\su}^F_3)\\
&=-{{\mathtt{v}}}^{{\text{in}}}_+-({{\mathpzc{Q}}}-2){\su}_{1;+}
-({\su}^{{\text{in}}}_{-1, 1}-{\su}^{{\text{in}}}_{0, 1})+(1-{z}){\su}_{0, 1},
\label{uF_k_sq_R}
\end{eqn}
on ${\mathscr{A}},$
where 
${\mathpzc{Q}}$ is defined by \eqref{dHelmholtzF_sq} (see also \cite{Bls1}).

With ${\su}^F_{3}={\mathpzc{V}}_{{\mathtt{N}}} {\su}^F_{2}$, i.e.
\begin{eqn}
{\su}^F_{3}={\su}^F_{2}{\mathpzc{V}}_{{\mathtt{N}}}, \quad{\mathpzc{V}}_{{\mathtt{N}}}=\frac{{\mathtt{V}}_{{\mathtt{N}}-3}({\vartheta})}{{\mathtt{V}}_{{\mathtt{N}}-2}({\vartheta})},
\label{u3Fu2F_sq_R}
\end{eqn}
the pair of above equations \eqref{uF_k_sq_R} can be expressed as
\begin{eqn}
{{\su}_{2;+}}-{{\su}_{1;+}}&={\su}_{2;-}^{{\text{in}}}-{\su}_{1;-}^{{\text{in}}}-({\mathpzc{Q}}-1-{\mathpzc{V}}_{{\mathtt{N}}}){\su}_2^F, \\
-({{\mathpzc{Q}}}-1) {\su}_{1;+}&=-{{\su}_{2;+}}+({\su}^{{\text{in}}}_{-1, 1}-{\su}^{{\text{in}}}_{{0}, 1})+({z}-1){\su}_{0, 1}.
\label{u2Fu1Fpair_sq_R}
\end{eqn}
With $({\mathpzc{Q}}-1){{\su}^{{\text{in}}}_{1;-}}={{\su}^{{\text{in}}}_{2;-}}-({\su}^{{\text{in}}}_{-1, 1}-{z}{\su}^{{\text{in}}}_{{0}, 1})$, the Wiener--Hopf equation for ${\su}_{2}^F$ (as the only unknown in \eqref{bulk_k_sq_R} after substitution of \eqref{u3Fu2F_sq_R}) is found to be
\beqan
{{\mathpzc{L}}}{{\su}_{2;+}}+{\su}_{2-}&=&-(1-{{\mathpzc{L}}})({{\su}_{2;-}^{{\text{in}}}}+{\mathpzc{H}}^{-1}
(1-{z}){\su}^{{t}}_{0, 1}), \label{WHKeq_sq_R}\\
\text{where }
{{\mathpzc{L}}}&=&1+\frac{{\mathpzc{H}}}{({{\mathpzc{Q}}}-1)({\mathpzc{Q}}-1-{\mathpzc{V}}_{{\mathtt{N}}})},
\label{LNgen_k_sq_R}
\eeqan
on ${\mathscr{A}}.$
Note that $(({{\mathpzc{Q}}}-1)({\mathpzc{Q}}-1-{\mathpzc{V}}_{{\mathtt{N}}}))^{-1}=({{\mathpzc{Q}}}-2)^{-1}({{\mathpzc{L}}}-1)$ and ${\su}_{1;+}$ is given by \eqref{u2Fu1Fpair_sq_R}.
Indeed,
\begin{eqn}
{{\mathpzc{L}}}&=\frac{{\mathtt{U}}_{{\mathtt{N}}-1}({\vartheta})}{({{\mathpzc{Q}}}-1){\mathtt{U}}_{{\mathtt{N}}-2}({\vartheta})}=\frac{{\mathscr{N}}}{{\mathscr{D}}}=\frac{{\mathscr{N}}}{({\mathpzc{Q}}-1)\mathring{{\mathscr{D}}}}.
\label{Lk_sq_R}
\end{eqn}
\begin{remark}
Notice that ${\mathscr{N}}$ and ${\mathscr{D}}$ in this case \eqref{Lk_sq_R} admit an `extra common' factor ${\mathpzc{h}}^2$ due to the presence of a wave mode for ${\upomega}\in[0,2]$ for which the step is transparent.
\label{remcasee}\end{remark}

\subsubsection{Case: (f)}
\label{crack_sq_XR}
{In this case} the particles at ${\mathtt{x}}\in\mathbb{Z}, {\mathtt{y}}={{\mathtt{N}}}$ are fixed while those at ${\mathtt{x}}\ge0, {\mathtt{y}}=1$ as well as at ${\mathtt{x}}<0, {\mathtt{y}}=1$ {belong} to a free boundary, as shown in Fig. \ref{stepboundarywaveguides_sq}(f). Thus, ${\mathfrak{S}\hspace{-.4ex}}{\mathbin{\substack{\bullet\\\circ\circ}}}$ is the waveguide constructed by a junction between ${\mathfrak{S}{\sbfrac{\bullet}{\circ}}}$ and ${\mathfrak{S}{\sbfrac{\bullet}{\circ}}}$ of widths ${\mathtt{N}}-1$ and ${\mathtt{N}}$, respectively.
For this choice of upper boundary \eqref{bc_sq_X}, using \eqref{U_k_sq_X}, 
\begin{eqn}
{\su}_{{\mathtt{y}}}^F ={\su}^F_2{\Lambda}_{{\mathtt{y}}}({\mathtt{N}}-1,{\bullet};1), {{\mathtt{y}}}\in\mathbb{Z}_2^{{\mathtt{N}}}, \label{bulk_k_sq_XR}
\end{eqn}
following the derivation for \S\ref{crack_sq_R}, 
\begin{eqn}
{\su}^F_{3}={\su}^F_{2}{\mathpzc{V}}_{{\mathtt{N}}}, \quad{\mathpzc{V}}_{{\mathtt{N}}}=\frac{{\mathtt{U}}_{{\mathtt{N}}-3}({\vartheta})}{{\mathtt{U}}_{{\mathtt{N}}-2}({\vartheta})},
\label{u2Fu1F_sq_XR}
\end{eqn}
(similar to \eqref{u2Fu1F_sq_X}). Hence \eqref{WHKeq_sq_R} holds for ${\su}_{2}^F$ (as the only unknown in \eqref{bulk_k_sq_XR} after substitution of \eqref{u2Fu1F_sq_XR}) with the kernel \eqref{LNgen_k_sq_R} simplified to the form
\begin{eqn}
{{\mathpzc{L}}}&=\frac{{\mathtt{V}}_{{\mathtt{N}}}({\vartheta})}{({{\mathpzc{Q}}}-1){\mathtt{V}}_{{\mathtt{N}}-1}({\vartheta})}=\frac{{\mathscr{N}}}{{\mathscr{D}}}=\frac{{\mathscr{N}}}{({\mathpzc{Q}}-1)\mathring{{\mathscr{D}}}}.
\label{Lk_sq_XR}
\end{eqn}

\subsubsection{Case: (g)}
\label{slit_sq_X}
{In this case} the particles at ${\mathtt{x}}\in\mathbb{Z}, {\mathtt{y}}={{\mathtt{N}}}+1$, ${\mathtt{x}}\ge0, {\mathtt{y}}=0$ as well as ${\mathtt{x}}<0, {\mathtt{y}}=1$ are fixed, as shown in Fig. \ref{stepboundarywaveguides_sq}(g). Thus, ${\mathfrak{S}\hspace{-.4ex}}{\mathbin{\substack{\bullet\\\bullet\bullet}}}$ is the waveguide constructed by a junction between ${\mathfrak{S}{\sbfrac{\bullet}{\bullet}}}$ and ${\mathfrak{S}{\sbfrac{\bullet}{\bullet}}}$ of widths ${\mathtt{N}}-1$ and ${\mathtt{N}}$, respectively.
For this choice of upper boundary \eqref{bc_sq_X}, \eqref{bulk_k_sq_X} is the ansatz for solution. As an analogue of \eqref{dHelmholtz_sq} for ${\mathtt{y}}=1,$
\begin{eqn}
-{\upomega}^2{\su}_{{{\mathtt{x}}}, 1}&=({\su}_{{{\mathtt{x}}}-1, 1}+{\su}_{{{\mathtt{x}}}+1, 1}+{\su}_{{{\mathtt{x}}}, 2}-4{\su}_{{{\mathtt{x}}}, 1}){\mathit{H}}({\mathtt{x}})
+{\upomega}^2{\su}^{{\text{in}}}_{{{\mathtt{x}}}, 1}{\mathit{H}}(-1-{\mathtt{x}}).\label{dimnewtoneqslitl_sq_X}
\end{eqn}
By an application of the Fourier transform \eqref{unpm}, 
it follows that
\begin{eqn}
-{\upomega}^2{\su}_{1;-}&={\su}_{-1, 1}-{{z}} {\su}_{0, 1}+({z}+{z}^{-1}-4+{\upomega}^2){\su}_{1;+}
+{\su}_{2;+}+{\upomega}^2{\su}^{{\text{in}}}_{1;-},
\label{uF_c_sq_X}
\end{eqn}
on ${\mathscr{A}}.$
After the substitution of \eqref{bulk_k_sq_X}, the expression of ${\su}^F_{2}$ in terms of ${\su}^F_{1}$ is obtained as \eqref{u2Fu1F_sq_X}.
In \eqref{uF_c_sq_X}, using 
the expression of
${\mathpzc{Q}}$ \eqref{dHelmholtzF_sq},
\beqans
{{\mathpzc{W}}}({{z}})+{{\mathpzc{Q}}}({{z}}){\su}_{1;-}({{z}})&=&{{\mathpzc{Q}}}({{z}}){\su}^F_1({{z}})-{\su}_{2{;}+}({{z}}), 
\label{sliteq_sq_X}\\
{\su}_{1; -}({{z}})&=&-\su^{{\text{in}}}_{1;-}, 
\label{u1n_X}\\
\text{where }{{\mathpzc{W}}}({{z}})&{:=}&{\su}_{-1, 1}-{{z}} {\su}_{0, 1}, {{z}}\in\mathbb{C}\label{q2beta_X}.
\eeqans{u0_X}
By {a rearrangement of} 
\eqref{sliteq_sq_X}, {using} \eqref{u2Fu1F_sq_X} and \eqref{unpm}, it follows that ${\su}^F_{2}$ (as the only unknown in \eqref{bulk_k_sq_X} after substitution of \eqref{u2Fu1F_sq_X}) satisfies the Wiener--Hopf equation
\beqans
{{\mathpzc{L}}}{\su}^F_{2;+}+{\su}^F_{2;-}&=&(1-{{\mathpzc{L}}})({{\mathpzc{W}}}-{{\mathpzc{Q}}}{\su}^{{\text{in}}}_{1;-}) \text{ on }{\mathscr{A}},\label{WHCeq_sq_X}\\
\text{where }
{{\mathpzc{L}}}&=&1-\frac{{\mathpzc{V}}_{{\mathtt{N}}}}{{\mathpzc{Q}}}. \label{WHkernel_sq_X}
\eeqans{WHeqslitfull_sq_X}
In \eqref{WHCeq_sq_X}, further, it can be shown that
\begin{eqn}
{{\mathpzc{L}}}&=\frac{1}{{\mathpzc{Q}}}\frac{{\mathtt{U}}_{{\mathtt{N}}}}{{\mathtt{U}}_{{\mathtt{N}}-1}}=\frac{{\mathscr{N}}}{{\mathscr{D}}}=\frac{{\mathscr{N}}}{{\mathpzc{Q}}\mathring{{\mathscr{D}}}}.
\label{Lc_sq_X}
\end{eqn}

\subsubsection{Case: (h)}
\label{slit_sq_RX}
{In this case} particles at ${\mathtt{x}}\in\mathbb{Z}, {\mathtt{y}}={{\mathtt{N}}}$ {belong} to a free boundary while those at ${\mathtt{x}}\ge0, {\mathtt{y}}=0$ as well as at ${\mathtt{x}}<0, {\mathtt{y}}=1$ are fixed, as shown in Fig. \ref{stepboundarywaveguides_sq}(h). Thus, ${\mathfrak{S}\hspace{-.4ex}}{\mathbin{\substack{\circ\\\bullet\bullet}}}$ is the waveguide constructed by a junction between ${\mathfrak{S}{\sbfrac{\circ}{\bullet}}}$ and ${\mathfrak{S}{\sbfrac{\circ}{\bullet}}}$ of widths ${\mathtt{N}}-1$ and ${\mathtt{N}}$, respectively.
For this choice of upper boundary \eqref{bc_sq_R}, the ansatz for solution is given by \eqref{bulk_k_sq_RX} (and \eqref{U_sq_RX}). Following \S\ref{slit_sq_X}, the expression of ${\su}^F_{1}$ in terms of ${\su}^F_{2}$ is found to be \eqref{u2Fu1F_sq_RX}, 
leading to \eqref{sliteq_sq_X} and \eqref{q2beta_X}, and eventually Wiener--Hopf equation \eqref{WHCeq_sq_X} (for the only unknown ${\su}^F_{2}$ in \eqref{bulk_k_sq_RX} after substitution of \eqref{u2Fu1F_sq_RX}) where
\begin{eqn}
{{\mathpzc{L}}}&=\frac{1}{{\mathpzc{Q}}}\frac{{\mathtt{V}}_{{\mathtt{N}}}}{{\mathtt{V}}_{{\mathtt{N}}-1}}=\frac{{\mathscr{N}}}{{\mathscr{D}}}=\frac{{\mathscr{N}}}{{\mathpzc{Q}}\mathring{{\mathscr{D}}}}.
\label{Lc_sq_RX}
\end{eqn}

\subsubsection{Case: (i)}
\label{crack_sq_RX2}
{In this case} the particles at ${\mathtt{x}}\in\mathbb{Z}, {\mathtt{y}}={{\mathtt{N}}}$, ${\mathtt{x}}\ge0, {\mathtt{y}}=1$ are attached to a fixed boundary at ${\mathtt{y}}=0$, while those at ${\mathtt{x}}<0, {\mathtt{y}}=2$ {belong} to a free boundary, as shown in Fig. \ref{stepboundarywaveguides_sq}(i). Thus, ${\mathfrak{S}\hspace{-.4ex}}{\mathbin{\substack{\circ\\\circ\overline{\bullet}}}}$ is the waveguide constructed by a junction between ${\mathfrak{S}{\sbfrac{\circ}{\circ}}}$ and ${\mathfrak{S}{\sbfrac{\circ}{\bullet}}}$ of widths ${\mathtt{N}}-1$ and ${\mathtt{N}}$, respectively.
For this choice of upper boundary \eqref{bc_sq_R}, using \eqref{U_sq_RX},
the ansatz for solution is \begin{eqn}{\su}_{{\mathtt{y}}}^F ={\su}^F_2{\Lambda}_{{\mathtt{y}}}({\mathtt{N}}-1,{\circ};1), {\mathtt{y}}\in\mathbb{Z}_2^{{\mathtt{N}}}.\label{bulk_k_sq_RX2}\end{eqn}
Equivalent to \eqref{dHelmholtz_sq} for ${\mathtt{y}}=2, 1,$
\beqans
(-{\upomega}^2{\su}_{{{\mathtt{x}}}, 2}-{{{b}^2}}{{{\mathit{v}}}^{{t}}}_{{\mathtt{x}}})=-({\su}_{{\mathtt{x}}, 1}^{{\text{in}}}-{\su}_{{\mathtt{x}}, 2}^{{\text{in}}})
+({\su}_{{{\mathtt{x}}}+1, 2}+{\su}_{{{\mathtt{x}}}-1, 2}+{\su}_{{{\mathtt{x}}}, 3}-3{\su}_{{{\mathtt{x}}}, 2}), {\mathtt{x}}\in\mathbb{Z},\label{dimnewtoneqcracku_sq_RX2}\\
(-{\upomega}^2{\su}_{{{\mathtt{x}}}, 1}+{{{b}^2}}{{{\mathit{v}}}^{{t}}}_{{\mathtt{x}}})=({\su}_{{\mathtt{x}}, 1}^{{\text{in}}}-{\su}_{{\mathtt{x}}, 2}^{{\text{in}}}-{\su}_{{\mathtt{x}}, 0}^{{\text{in}}}
+({\su}_{{{\mathtt{x}}}-1, 1}-{\su}_{{{\mathtt{x}}}, 1}){\mathit{H}}({\mathtt{x}}-1)+({\su}_{{{\mathtt{x}}}+1, 1}-2{\su}_{{{\mathtt{x}}}, 1}){\mathit{H}}({\mathtt{x}})\notag\\
-({\su}^{{\text{in}}}_{{{\mathtt{x}}}-1, 1}-{\su}^{{\text{in}}}_{{{\mathtt{x}}}, 1}){\updelta}_{{\mathtt{x}},0}, {\mathtt{x}}\ge0,\label{dimnewtoneqcrackl_sq_RX2}
\eeqans{dimnewtoneqcrack_sq_RX2}
where ${{{\mathit{v}}}^{{t}}}_{{\mathtt{x}}}$ is the total force that acts on the particle $({{\mathtt{x}}}, 2)$, 
i.e.
\begin{eqn}
{{{b}^2}}{{{\mathit{v}}}^{{t}}}^F=-({{\su}_{{\mathtt{x}}, 2}^{{t}}}-{{\su}_{{\mathtt{x}}, 1}^{{t}}}){\mathit{H}}({\mathtt{x}}), \quad {{{\mathtt{v}}}^{{\text{in}}}_{{\mathtt{x}}}}{:=}{{\su}_{{\mathtt{x}}, 2}^{{\text{in}}}}-{{\su}_{{\mathtt{x}}, 1}^{{\text{in}}}}, {\mathtt{x}}\in\mathbb{Z},
\label{vinc_def_RX2}
\end{eqn}
(similar to \eqref{vinc_def_X}). By an application of the Fourier transform \eqref{unpm}, 
it follows that
\begin{eqn}
-{{\mathtt{v}}}_+-{{\mathtt{v}}}^{{\text{in}}}_+={{{b}^2}}{{{\mathit{v}}}^{{t}}}^F
&=-{{{\mathtt{v}}}^{{\text{in}}}}^F+(({{\mathpzc{Q}}}-1) {\su}^F_2-{\su}^F_3)\\
&=-{{\mathtt{v}}}^{{\text{in}}}_+-{\su}^{{\text{in}}}_{0;+}-({{\mathpzc{Q}}}-1){\su}_{1;+}-({\su}^{{\text{in}}}_{-1, 1}-{\su}^{{\text{in}}}_{0, 1})+(1-{z}){\su}_{0, 1},
\label{uF_k_sq_RX2}
\end{eqn}
where 
${\mathpzc{Q}}$ is defined by \eqref{dHelmholtzF_sq} (see also \cite{Bls1}).
With ${\su}^F_{3}={\mathpzc{V}}_{{\mathtt{N}}} {\su}^F_{2}$, i.e.
\begin{eqn}
{\su}^F_{3}={\su}^F_{2}{\mathpzc{V}}_{{\mathtt{N}}}, \quad{\mathpzc{V}}_{{\mathtt{N}}}=\frac{{\mathtt{V}}_{{\mathtt{N}}-3}({\vartheta})}{{\mathtt{V}}_{{\mathtt{N}}-2}({\vartheta})},
\label{u3Fu2F_sq_RX2}
\end{eqn}
the pair of above equations \eqref{uF_k_sq_RX2} can be expressed as
\begin{eqn}
{{\su}_{2;+}}-{{\su}_{1;+}}&={\su}_{2;-}^{{\text{in}}}-{\su}_{1;-}^{{\text{in}}}-({\mathpzc{Q}}-1-{\mathpzc{V}}_{{\mathtt{N}}}){\su}_2^F, \\
-{{\mathpzc{Q}}}{\su}_{1;+}&=-{{\su}_{2;+}}+{\su}_{0;+}^{{\text{in}}}+({\su}^{{\text{in}}}_{-1, 1}-{\su}^{{\text{in}}}_{{0}, 1})+({z}-1){\su}_{0, 1}.
\label{u2Fu1Fpair_sq_RX2}
\end{eqn}
With ${\mathpzc{Q}}{{\su}^{{\text{in}}}_{1;-}}={{\su}^{{\text{in}}}_{2;-}}+{{\su}^{{\text{in}}}_{0;-}}-({\su}^{{\text{in}}}_{-1, 1}-{z}{\su}^{{\text{in}}}_{{0}, 1})$, the Wiener--Hopf equation for ${\su}_{2}^F$ (as the only unknown in \eqref{bulk_k_sq_RX2} after substitution of \eqref{u3Fu2F_sq_RX2}) is found to be (on ${\mathscr{A}}$)
\beqan
{{\mathpzc{L}}}{{\su}_{2;+}}+{\su}_{2-}&=&-(1-{{\mathpzc{L}}})({{\su}_{2;-}^{{\text{in}}}}+({\mathpzc{Q}}-1)^{-1}(1-{z}){\su}^{{t}}_{0, 1}), \label{WHKeq_sq_RX2}\\
\text{where }
{{\mathpzc{L}}}&=&1+\frac{{\mathpzc{Q}}-1}{{{\mathpzc{Q}}}({\mathpzc{Q}}-1-{\mathpzc{V}}_{{\mathtt{N}}})}.
\label{LNgen_k_sq_RX2}
\eeqan
Note that $({{\mathpzc{Q}}}({\mathpzc{Q}}-1-{\mathpzc{V}}_{{\mathtt{N}}}))^{-1}=({{\mathpzc{Q}}}-1)^{-1}({{\mathpzc{L}}}-1)$ and ${\su}_{1;+}$ is given by \eqref{u2Fu1Fpair_sq_RX2}.
Hence,
\begin{eqn}
{{\mathpzc{L}}}&=\frac{{\mathtt{V}}_{{\mathtt{N}}}({\vartheta})}{{\mathpzc{Q}}({{\mathpzc{Q}}}-2){\mathtt{U}}_{{\mathtt{N}}-2}({\vartheta})}=\frac{{\mathscr{N}}}{{\mathscr{D}}}=\frac{{\mathscr{N}}}{{\mathpzc{Q}}\mathring{{\mathscr{D}}}}.
\label{Lk_sq_RX2}
\end{eqn}

\subsubsection{Case: (j)}
\label{crack_sq_X2}
{In this case} the particles at ${\mathtt{x}}\in\mathbb{Z}, {\mathtt{y}}={{\mathtt{N}}}$ and at ${\mathtt{x}}\ge0, {\mathtt{y}}=0$ are fixed while those at ${\mathtt{x}}<0, {\mathtt{y}}=1$ {belong} to a free boundary, as shown in Fig. \ref{stepboundarywaveguides_sq}(j). Thus, ${\mathfrak{S}\hspace{-.4ex}}{\mathbin{\substack{{\bullet}\\{\circ}\overline{\bullet}}}}$ is the waveguide constructed by a junction between ${\mathfrak{S}{\sbfrac{\bullet}{\circ}}}$ and ${\mathfrak{S}{\sbfrac{\bullet}{\bullet}}}$ of widths ${\mathtt{N}}-1$ and ${\mathtt{N}}$, respectively.
For this choice of upper boundary \eqref{bc_sq_X}, using \eqref{U_k_sq_X}, 
\begin{eqn}
{\su}_{{\mathtt{y}}}^F ={\su}^F_2{\Lambda}_{{\mathtt{y}}}({\mathtt{N}}-1,{\bullet};1), {{\mathtt{y}}}\in\mathbb{Z}_2^{{\mathtt{N}}}, \label{bulk_k_sq_X2}
\end{eqn}
following the derivation for \S\ref{crack_sq_RX2}, 
\begin{eqn}
{\su}^F_{3}={\su}^F_{2}{\mathpzc{V}}_{{\mathtt{N}}}, \quad{\mathpzc{V}}_{{\mathtt{N}}}=\frac{{\mathtt{U}}_{{\mathtt{N}}-3}({\vartheta})}{{\mathtt{U}}_{{\mathtt{N}}-2}({\vartheta})},
\label{u2Fu1F_sq_X2}
\end{eqn}
(similar to \eqref{u2Fu1F_sq_X}). Hence \eqref{WHKeq_sq_R} holds for ${\su}_{2}^F$ (as the only unknown in \eqref{bulk_k_sq_X2} after substitution of \eqref{u2Fu1F_sq_X2}) with the kernel \eqref{LNgen_k_sq_R}
Hence,
\begin{eqn}
{{\mathpzc{L}}}&=
\frac{{\mathtt{U}}_{{\mathtt{N}}}({\vartheta})}{{{\mathpzc{Q}}}{\mathtt{V}}_{{\mathtt{N}}-1}({\vartheta})}=\frac{{\mathscr{N}}}{{\mathscr{D}}}=\frac{{\mathscr{N}}}{{\mathpzc{Q}}\mathring{{\mathscr{D}}}}.
\label{Lk_sq_X2}
\end{eqn}

\subsection{Incidence from the other portion of the lattice strip}
In order to construct the complete transmission properties, the alternative incidence from the portion behind the step
is also required to be analyzed. Suppose that ${{\mathfrak{s}}}={A}$ denotes incidence from the portion {\em ahead} of the step discontinuity while ${{\mathfrak{s}}}={B}$ denotes incidence from the portion {\em behind} the step discontinuity.
By inspection of the Fig. \ref{stepboundarywaveguides_sq}, as well as the analysis presented above for the incidence for ${{\mathfrak{s}}}={A}$, it suffices to consider the cases of
${\mathfrak{S}\hspace{-.4ex}}{\mathbin{\substack{{\bullet}\\ {\circ}{\bullet}}}}$, 
${\mathfrak{S}\hspace{-.4ex}}{\mathbin{\substack{\circ\\\bullet\circ}}}$, 
${\mathfrak{S}\hspace{-.4ex}}{\mathbin{\substack{\circ\\\circ\circ}}}$, ${\mathfrak{S}\hspace{-.4ex}}{\mathbin{\substack{\bullet\\\bullet\bullet}}}$, 
and ${\mathfrak{S}\hspace{-.4ex}}{\mathbin{\substack{\circ\\\circ\overline{\bullet}}}}$, since the Wiener--Hopf equation retains the 
same form in ${\mathfrak{S}\hspace{-.4ex}}{\mathbin{\substack{\circ\\\circ\bullet}}}$, ${\mathfrak{S}\hspace{-.4ex}}{\mathbin{\substack{\bullet\\\bullet\circ}}}$, 
${\mathfrak{S}\hspace{-.4ex}}{\mathbin{\substack{\bullet\\\circ\circ}}}$, 
${\mathfrak{S}\hspace{-.4ex}}{\mathbin{\substack{\circ\\\bullet\bullet}}}$, 
and 
${\mathfrak{S}\hspace{-.4ex}}{\mathbin{\substack{{\bullet}\\ {\circ}\overline{\bullet}}}}$, 
respectively; in particular, the right hand side in the respective cases is the same.

\subsubsection{Case: (a), (b)}
In the case of incidence from the other direction ${{\mathfrak{s}}}={B},$, due to the presence of {normal mode factor} in \eqref{uinc_sq}, ${{\su}^{{\text{in}}}_{\cdot, 1}}$ satisfies the free boundary condition \eqref{bc_sq_R} (choosing $+$ sign), and ${{\su}^{{\text{in}}}_{\cdot, {\mathtt{N}}+1}}=0$. 
In place of \eqref{dHelmholtz_sq} for ${\mathtt{y}}=1,$
$-{\upomega}^2{\su}_{{{\mathtt{x}}}, 1}+{\su}_{{{\mathtt{x}}}, 1}{\mathit{H}}({\mathtt{x}})+{\su}^{{\text{in}}}_{{{\mathtt{x}}}, 1}{\mathit{H}}({\mathtt{x}})={\su}_{{{\mathtt{x}}}+1, 1}+{\su}_{{{\mathtt{x}}}-1, 1}+{\su}_{{{\mathtt{x}}}, 2}-3{\su}_{{{\mathtt{x}}}, 1}.$
By an application of the Fourier transform \eqref{unpm}, 
it follows that
\begin{eqn}
(({{\mathpzc{Q}}}-1) {\su}^F_1-{\su}^F_2)+{\su}_{1;+}=-{\su}^{{\text{in}}}_{1;+}.
\label{uF_k_sq_X_altinc}
\end{eqn}
On comparison of \eqref{uF_k_sq_X} and \eqref{uF_k_sq_X_altinc}, the Wiener--Hopf equation (compare with \eqref{WHKeq_sq_X}) is found to be
\begin{eqn}
{{\mathpzc{L}}}{\su}_{1;+}+{\su}_{1;-}&=(1-{\mathpzc{L}})\su^{{\text{in}}}_{1;+}. 
\label{WHKeq_sq_X_altinc}
\end{eqn}

\subsubsection{Case: (c), (d)}
In the case of incidence from the other direction ${{\mathfrak{s}}}={B},$ due to the presence of {normal mode factor} in \eqref{uinc_sq}, as an analogue of \eqref{dHelmholtz_sq} for ${\mathtt{y}}=1,$
$-{\upomega}^2{\su}_{{{\mathtt{x}}}, 1}=({\su}_{{{\mathtt{x}}}-1, 1}+{\su}_{{{\mathtt{x}}}+1, 1}+{\su}_{{{\mathtt{x}}}, 2}-3{\su}_{{{\mathtt{x}}}, 1}){\mathit{H}}({\mathtt{x}})+{\su}^{{\text{in}}}_{{{\mathtt{x}}}, 2}{\mathit{H}}({\mathtt{x}}).$\label{dimnewtoneqslitl_sq_R_altinc}
The resulting equation 
\begin{eqn}
{{\mathpzc{W}}}+{\su}^{{\text{in}}}_{2;+}=({\mathpzc{Q}}-1){\su}^F_{1}-{\su}_{2;+}, 
\label{sliteq_sq_R_altinc}
\end{eqn}
can be compared with \eqref{sliteq_sq_R}. Thus, a Wiener--Hopf equation can be found (compare with \eqref{WHCeq_sq_R})
\begin{eqn}
{{\mathpzc{L}}}{\su}_{2;+}+{\su}_{2;-}&=(1-{{\mathpzc{L}}})({{\mathpzc{W}}}
+{\su}^{{\text{in}}}_{2;+}) \text{ on }{\mathscr{A}}.
\label{WHCeq_sq_R_altinc}
\end{eqn}

\subsubsection{Case: (e), (f)}
In the case of incidence from the other direction ${{\mathfrak{s}}}={B},$ due to the presence of {normal mode factor} in \eqref{uinc_sq}, using the equation corresponding to \eqref{dHelmholtz_sq} for ${\mathtt{y}}=2, 1,$
leads to
\beqans
(-{\upomega}^2{\su}_{{{\mathtt{x}}}, 2}-{{{b}^2}}{{{\mathit{v}}}^{{t}}}_{{\mathtt{x}}})&=&{\su}_{{{\mathtt{x}}}+1, 2}+{\su}_{{{\mathtt{x}}}-1, 2}+{\su}_{{{\mathtt{x}}}, 3}-3{\su}_{{{\mathtt{x}}}, 2},\label{dimnewtoneqcracku_sq_R_altinc}\\
(-{\upomega}^2{\su}_{{{\mathtt{x}}}, 1}+{{{b}^2}}{{{\mathit{v}}}^{{t}}}_{{\mathtt{x}}})&=&({\su}_{{{\mathtt{x}}}-1, 1}-{\su}_{{{\mathtt{x}}}, 1}){\mathit{H}}({\mathtt{x}}-1)
+({\su}_{{{\mathtt{x}}}+1, 1}-{\su}_{{{\mathtt{x}}}, 1}){\mathit{H}}({\mathtt{x}}).
\label{dimnewtoneqcrackl_sq_R_altinc}
\eeqans{dimnewtoneqcrack_sq_R_altinc}
Further, the equations \eqref{dimnewtoneqcrack_sq_R_altinc} can be expressed as
$-{{\mathtt{v}}}_{+}-{{\mathtt{v}}}^{{\text{in}}}_{+}=(({{\mathpzc{Q}}}-1) {\su}^F_2-{\su}^F_3)=-({{\mathpzc{Q}}}-2){\su}^F_{1}-({z}-1){\su}_{0, 1}.$
Following closely the manipulations stated earlier for 
${{\mathfrak{s}}}={A},$
\begin{eqn}
{{\su}_{2;+}}-{{\su}_{1;+}}&=-{{\su}^{{\text{in}}}_{2;+}}-({\mathpzc{Q}}-1-{\mathpzc{V}}_{{\mathtt{N}}}){\su}_2^F, \\
-({{\mathpzc{Q}}}-1) {\su}_{1;+}+{{\su}^{{\text{in}}}_{2;+}}&=-{{\su}_{2;+}}+({z}-1){\su}_{0, 1}, 
\label{u2Fu1Fpair_sq_R_altinc}
\end{eqn}
Using ${{\su}_{1;-}}={{\su}^{{\text{in}}}_{1;-}}=0$, a scalar Wiener--Hopf equation can be constructed (compare with \eqref{WHKeq_sq_R})
\begin{eqn}
{{\mathpzc{L}}}{{\su}_{2;+}}+{\su}_{2-}
&=(1-{{\mathpzc{L}}})({{\su}_{2;+}^{{\text{in}}}}-{\mathpzc{H}}^{-1}(1-{z}){\su}_{0, 1})).
\label{WHKeq_sq_R_altinc}
\end{eqn}

\subsubsection{Case: (g), (h)}
In the case of incidence from the other direction, ${{\mathfrak{s}}}={B},$ due to the presence of {normal mode factor} in \eqref{uinc_sq}, as an analogue of \eqref{dHelmholtz_sq} for ${\mathtt{y}}=1,$
\beqan
-{\upomega}^2{\su}_{{{\mathtt{x}}}, 1}&=&({\su}_{{{\mathtt{x}}}-1, 1}+{\su}_{{{\mathtt{x}}}+1, 1}+{\su}_{{{\mathtt{x}}}, 2}-4{\su}_{{{\mathtt{x}}}, 1}){\mathit{H}}({\mathtt{x}})
+({\su}^{{\text{in}}}_{{{\mathtt{x}}}-1, 1}+{\su}^{{\text{in}}}_{{{\mathtt{x}}}+1, 1}+{\su}^{{\text{in}}}_{{{\mathtt{x}}}, 2}-4{\su}^{{\text{in}}}_{{{\mathtt{x}}}, 1}){\mathit{H}}({\mathtt{x}}).
\eeqan
Thus,
$-{\upomega}^2{\su}_{1;-}={\su}_{-1, 1}-{{z}} {\su}_{0, 1}+({z}+{z}^{-1}-4+{\upomega}^2){\su}_{1;+}+{\su}_{2;+}+{\su}^{{\text{in}}}_{2;+},$
which leads to
\begin{eqn}
({\su}_{2;+}+{\su}_{2;-})-{\mathpzc{V}}_{{\mathtt{N}}}{\mathpzc{Q}}^{-1}{\su}_{2;+}&={\mathpzc{V}}_{{\mathtt{N}}}{\mathpzc{Q}}^{-1}({{\mathpzc{W}}}+{\su}^{{\text{in}}}_{2;+}).
\label{dimnewtoneqslitl_sq_X_altinc}
\end{eqn}
Thus, \eqref{dimnewtoneqslitl_sq_X_altinc} yields the scalar Wiener--Hopf equation (compare with \eqref{WHCeq_sq_X})
\begin{eqn}
{{\mathpzc{L}}}{\su}_{2;+}+{\su}_{2;-}&=(1-{{\mathpzc{L}}})({{\mathpzc{W}}}+{\su}^{{\text{in}}}_{2;+}) \text{ on }{\mathscr{A}}.
\label{WHCeq_sq_X_altinc}
\end{eqn}

\subsubsection{Case: (i), (j)}
In the case of incidence from the other direction, ${{\mathfrak{s}}}={B},$ due to the presence of {normal mode factor} in \eqref{uinc_sq}, in place of \eqref{dHelmholtz_sq} for ${\mathtt{y}}=2, 1,$
leads to
\beqans
(-{\upomega}^2{\su}_{{{\mathtt{x}}}, 2}-{{{b}^2}}{{{\mathit{v}}}^{{t}}}_{{\mathtt{x}}})&=&({\su}_{{{\mathtt{x}}}+1, 2}+{\su}_{{{\mathtt{x}}}-1, 2}+{\su}_{{{\mathtt{x}}}, 3}-3{\su}_{{{\mathtt{x}}}, 2}),\label{dimnewtoneqcracku_sq_RX2_altinc}\\
(-{\upomega}^2{\su}_{{{\mathtt{x}}}, 1}+{{{b}^2}}{{{\mathit{v}}}^{{t}}}_{{\mathtt{x}}})&=&({\su}_{{{\mathtt{x}}}-1, 1}-{\su}_{{{\mathtt{x}}}, 1}){\mathit{H}}({\mathtt{x}}-1)
+({\su}_{{{\mathtt{x}}}+1, 1}+0-2{\su}_{{{\mathtt{x}}}, 1}){\mathit{H}}({\mathtt{x}}).
\label{dimnewtoneqcrackl_sq_RX2_altinc}
\eeqans{dimnewtoneqcrack_sq_RX2_altinc}
Further, the equations \eqref{dimnewtoneqcracku_sq_RX2_altinc} can be expressed as
$-{{\mathtt{v}}}_{+}-{{\mathtt{v}}}^{{\text{in}}}_{+}=(({{\mathpzc{Q}}}-1) {\su}^F_2-{\su}^F_3)=-({{\mathpzc{Q}}}-1){\su}_{1;+}+(1-{z}){\su}_{0, 1}.$
Following the manipulations stated earlier for 
${{\mathfrak{s}}}={A},$
\begin{eqn}
{{\su}_{2;+}}-{{\su}_{1;+}}&=-{{\su}_{2;+}^{{\text{in}}}}-({\mathpzc{Q}}-1-{\mathpzc{V}}_{{\mathtt{N}}}){\su}_2^F, \\
-{{\mathpzc{Q}}}{\su}_{1;+}&=-{{\su}^{{\text{in}}}_{2;+}}-{{\su}_{2;+}}+({z}-1){\su}_{0, 1}, 
\label{u2Fu1Fpair_sq_RX2_altinc}
\end{eqn}
Using ${{\su}_{1;-}}={{\su}^{{\text{in}}}_{1;-}}=0$, a scalar Wiener--Hopf equation can be constructed (compare with \eqref{WHKeq_sq_RX2})
\begin{eqn}
{{\mathpzc{L}}}{{\su}_{2;+}}+{\su}_{2-}&=(1-{{\mathpzc{L}}})({{\su}_{2;+}^{{\text{in}}}}-({\mathpzc{Q}}-1)^{-1}(1-{z}){\su}_{0, 1})).
\label{WHKeq_sq_RX2_altinc}
\end{eqn}

This concludes the analysis and discussion of all the cases illustrated in Fig. \ref{stepboundarywaveguides_sq} from the point of view of incidence from either side of the step discontinuity, i.e., ${{\mathfrak{s}}}={A}$ or ${B}$. 
It needs to be emphasized again that the Wiener--Hopf kernel ${{\mathpzc{L}}}$ remains the same between \eqref{WHKeq_sq_X} (resp. \eqref{WHCeq_sq_R}, \eqref{WHKeq_sq_R}, \eqref{WHCeq_sq_X}, and \eqref{WHKeq_sq_RX2}) and \eqref{WHKeq_sq_X_altinc} (resp. \eqref{WHCeq_sq_R_altinc}, \eqref{WHKeq_sq_R_altinc}, \eqref{WHCeq_sq_X_altinc}, and \eqref{WHKeq_sq_RX2_altinc}) (few extra details have been also provided in the supplementary \S1).

\subsection{Application of Wiener--Hopf method}\label{crack_sq_WHfac}
\label{sq_WHfac}
The choice of notation in the last part of the equations for ${{\mathpzc{L}}}$, providing its expression, represents that the denominator of ${{\mathpzc{L}}}$ contains dispersion relations for different modes behind the step discontinuity, while the numerator of ${{\mathpzc{L}}}$ contains dispersion relations for different modes ahead. Recall \ref{app_recallbifwavesq_Chebdef}, in particular, the relations \eqref{disp_sq_X}, \eqref{disp_sq_R}, and \eqref{disp_sq_XR} in this context. 
In each case of the ten lattice waveguides considered above, the Wiener--Hopf kernel ${{\mathpzc{L}}}$ is a ratio of a polynomial ${\mathscr{N}}$ (of ${z}$) in the numerator and another polynomial ${\mathscr{D}}$ in the denominator. Further, these polynomials admit simple quadratic factors as described in Appendix B.1 of \cite{Bls5k_tube} or \S3.4 of \cite{Bls9s}.
Evidently, analogous details corresponding to the Wiener--Hopf factorization hold for the square lattice strips; for instance, ${\mathfrak{S}\hspace{-.4ex}}{\mathbin{\substack{{\bullet}\\ {\circ}{\bullet}}}}$ using \eqref{Lk_sq_X} (see Fig. \ref{stepboundarywaveguides_sq}(b)), and also for the other boundary conditions.
All ten cases are also summarized,
also as a consequence of the identities involving Chebyshev polynomials \cite{Chebyshev00,MasonHand}, 
in Table \ref{stripbc_sq_ND} (with ${{z}}_{{{F}}}(\theta){:=}{\frac{1}{2}}(2+4\sin^2{\frac{1}{2}}\theta-{\upomega}^2-\sqrt{(2+4\sin^2{\frac{1}{2}}\theta-{\upomega}^2)^2-4})$).\footnote{Evidently, $({\mathpzc{h}}^2+2)$ in Table \ref{stripbc_sq_ND} can be also written as ${{F}}({{z}}; {{z}}_{{F}}({\frac{1}{2}}\pi))$; also $({\mathpzc{h}}^2+1)$ can be written as ${{F}}({{z}}; {{z}}_{{F}}(\frac{1}{3}\pi))$.}

\begin{table}[t]
\caption{{Wiener--Hopf kernel ${\mathscr{N}}/{\mathscr{D}}$ for the square lattice waveguides with a step on boundary (refer Fig. \ref{stepboundarywaveguides_sq})}}
\begin{center}
\begin{tabular}{|c|c|c|l|l|l|c|c|}
\hline
No.&strip&${\mathpzc{L}}$&${\mathscr{N}}$&${\mathscr{D}}$&W-H Eq&$\max N^{{A}}$&$\max N^{{B}}$\\
\hline
\hline
(a) &${\mathfrak{S}\hspace{-.4ex}}{\mathbin{\substack{{\bullet}\\ {\circ}{\bullet}}}}$&\eqref{Lk_sq_X}&$\prod\limits_{j=1}^{{\mathtt{N}}}{{F}}({{z}}; {{z}}_{{F}}(\frac{j}{{\mathtt{N}}+1}\pi))$&$\prod\limits_{j=1}^{{\mathtt{N}}}{{F}}({{z}}; {{z}}_{{F}}(\frac{j-{\frac{1}{2}}}{{\mathtt{N}}+{\frac{1}{2}}}\pi))$&\eqref{WHKeq_sq_X}, \eqref{WHKeq_sq_X_altinc}&${\mathtt{N}}$&${\mathtt{N}}$\\
(b) &${\mathfrak{S}\hspace{-.4ex}}{\mathbin{\substack{\circ\\\circ\bullet}}}$&\eqref{Lk_sq_RX}&$\prod\limits_{j=1}^{{\mathtt{N}}}{{F}}({{z}}; {{z}}_{{F}}(\frac{j-{\frac{1}{2}}}{{\mathtt{N}}+{\frac{1}{2}}}\pi))$&$\prod\limits_{j=0}^{{\mathtt{N}}-1}{{F}}({{z}}; {{z}}_{{F}}(\frac{j}{{\mathtt{N}}}\pi))$&\eqref{WHKeq_sq_X}, \eqref{WHKeq_sq_X_altinc}&${\mathtt{N}}$&${\mathtt{N}}$\\
\hline
(c) &${\mathfrak{S}\hspace{-.4ex}}{\mathbin{\substack{\circ\\\bullet\circ}}}$&\eqref{Lc_sq_R}&$\prod\limits_{j=0}^{{\mathtt{N}}-1}{{F}}({{z}}; {{z}}_{{F}}(\frac{j}{{\mathtt{N}}}\pi))$&$({\mathpzc{h}}^2+1)\prod\limits_{j=1}^{{\mathtt{N}}-1}{{F}}({{z}}; {{z}}_{{F}}(\frac{j-{\frac{1}{2}}}{{\mathtt{N}}-{\frac{1}{2}}}\pi))$&\eqref{WHCeq_sq_R}, \eqref{WHCeq_sq_R_altinc}&${\mathtt{N}}$&${\mathtt{N}}-1$\\
(d) &${\mathfrak{S}\hspace{-.4ex}}{\mathbin{\substack{\bullet\\\bullet\circ}}}$&\eqref{Lc_sq_XR}&$\prod\limits_{j=1}^{{\mathtt{N}}}{{F}}({{z}}; {{z}}_{{F}}(\frac{j-{\frac{1}{2}}}{{\mathtt{N}}+{\frac{1}{2}}}\pi))$&$({\mathpzc{h}}^2+1)\prod\limits_{j=1}^{{\mathtt{N}}-1}{{F}}({{z}}; {{z}}_{{F}}(\frac{j}{{\mathtt{N}}-{\frac{1}{2}}}\pi))$&\eqref{WHCeq_sq_R}, \eqref{WHCeq_sq_R_altinc}&${\mathtt{N}}$&${\mathtt{N}}-1$\\
\hline
(e) &${\mathfrak{S}\hspace{-.4ex}}{\mathbin{\substack{\circ\\\circ\circ}}}$&\eqref{Lk_sq_R}&$\prod\limits_{j=0}^{{\mathtt{N}}-1}{{F}}({{z}}; {{z}}_{{F}}(\frac{j}{{\mathtt{N}}}\pi))$&$({\mathpzc{h}}^2+1)\prod\limits_{j=0}^{{\mathtt{N}}-2}{{F}}({{z}}; {{z}}_{{F}}(\frac{j}{{\mathtt{N}}-1}\pi))$&\eqref{WHKeq_sq_R}, \eqref{WHKeq_sq_R_altinc}&${\mathtt{N}}$&${\mathtt{N}}-1$\\
(f) &${\mathfrak{S}\hspace{-.4ex}}{\mathbin{\substack{\bullet\\\circ\circ}}}$&\eqref{Lk_sq_XR}&$\prod\limits_{j=1}^{{\mathtt{N}}}{{F}}({{z}}; {{z}}_{{F}}(\frac{j-{\frac{1}{2}}}{{\mathtt{N}}+{\frac{1}{2}}}\pi))$&$({\mathpzc{h}}^2+1)\prod\limits_{j=0}^{{\mathtt{N}}-1}{{F}}({{z}}; {{z}}_{{F}}(\frac{j-{\frac{1}{2}}}{{\mathtt{N}}-{\frac{1}{2}}}\pi))$&\eqref{WHKeq_sq_R}, \eqref{WHKeq_sq_R_altinc}&${\mathtt{N}}$&${\mathtt{N}}-1$\\
\hline
(g) &${\mathfrak{S}\hspace{-.4ex}}{\mathbin{\substack{\bullet\\\bullet\bullet}}}$&\eqref{Lc_sq_X}&$\prod\limits_{j=1}^{{\mathtt{N}}}{{F}}({{z}}; {{z}}_{{F}}(\frac{j}{{\mathtt{N}}+1}\pi))$&$({\mathpzc{h}}^2+2)\prod\limits_{j=1}^{{\mathtt{N}}-1}{{F}}({{z}}; {{z}}_{{F}}(\frac{j}{{\mathtt{N}}}\pi))$&\eqref{WHCeq_sq_X}, \eqref{WHCeq_sq_X_altinc}&${\mathtt{N}}$&${\mathtt{N}}-1$\\
(h) &${\mathfrak{S}\hspace{-.4ex}}{\mathbin{\substack{\circ\\\bullet\bullet}}}$&\eqref{Lc_sq_RX}&$\prod\limits_{j=1}^{{\mathtt{N}}}{{F}}({{z}}; {{z}}_{{F}}(\frac{j-{\frac{1}{2}}}{{\mathtt{N}}+{\frac{1}{2}}}\pi))$&$({\mathpzc{h}}^2+2)\prod\limits_{j=1}^{{\mathtt{N}}-1}{{F}}({{z}}; {{z}}_{{F}}(\frac{j-{\frac{1}{2}}}{{\mathtt{N}}+{\frac{1}{2}}}\pi))$&\eqref{WHCeq_sq_X}, \eqref{WHCeq_sq_X_altinc}&${\mathtt{N}}$&${\mathtt{N}}-1$\\
\hline
(i) &${\mathfrak{S}\hspace{-.4ex}}{\mathbin{\substack{\circ\\\circ\overline{\bullet}}}}$&\eqref{Lk_sq_RX2}&$\prod\limits_{j=1}^{{\mathtt{N}}}{{F}}({{z}}; {{z}}_{{F}}(\frac{j-{\frac{1}{2}}}{{\mathtt{N}}-{\frac{1}{2}}}\pi))$&$({\mathpzc{h}}^2+2)\prod\limits_{j=0}^{{\mathtt{N}}-2}{{F}}({{z}}; {{z}}_{{F}}(\frac{j}{{\mathtt{N}}}\pi))$&\eqref{WHKeq_sq_RX2}, \eqref{WHKeq_sq_RX2_altinc}&${\mathtt{N}}$&${\mathtt{N}}-1$\\
(j) &${\mathfrak{S}\hspace{-.4ex}}{\mathbin{\substack{{\bullet}\\{\circ}\overline{\bullet}}}}$&\eqref{Lk_sq_X2}&$\prod\limits_{j=1}^{{\mathtt{N}}}{{F}}({{z}}; {{z}}_{{F}}(\frac{j}{{\mathtt{N}}+1}\pi))$&$({\mathpzc{h}}^2+2)\prod\limits_{j=1}^{{\mathtt{N}}-1}{{F}}({{z}}; {{z}}_{{F}}(\frac{j-{\frac{1}{2}}}{{\mathtt{N}}-{\frac{1}{2}}}\pi))$&\eqref{WHKeq_sq_RX2}, \eqref{WHKeq_sq_RX2_altinc}&${\mathtt{N}}$&${\mathtt{N}}-1$\\
\hline
\end{tabular}
\end{center}
\label{stripbc_sq_ND}
\end{table}

In general, the right hand side ${{\mathpzc{C}}}_{{A},{B}}$ in the Wiener--Hopf equations \eqref{WHKeq_sq_X} (resp. \eqref{WHCeq_sq_R}, \eqref{WHKeq_sq_R}, \eqref{WHCeq_sq_X}) and \eqref{WHKeq_sq_X_altinc} (resp. \eqref{WHCeq_sq_R_altinc}, \eqref{WHKeq_sq_R_altinc}, \eqref{WHCeq_sq_X_altinc}) can be together expressed as \begin{eqn}
{{\mathpzc{C}}}_{{A}}\delta_{{{\mathfrak{s}}},{A}}+{\mathpzc{C}}_{{B}}\delta_{{{\mathfrak{s}}},{B}}, 
\label{Cs_sq_def}
\end{eqn}
where ${\mathfrak{s}}={A}$ represents incidence from the right portion (ahead) and ${\mathfrak{s}}={B}$ denotes that from the left (behind) of the step discontinuity. It is understood as a useful notation that ${{\mathfrak{s}}}$ denotes any of the two cases of the incidence.
In the same context, it is also useful to define 
\begin{eqn}
\delta_{D+}({{z}})&{:=}\sum\nolimits_{n=0}^{+\infty}{{z}}^{-n}, |{{z}}|>1, 
\delta_{D-}({{z}}){:=}\sum\nolimits_{n=-\infty}^{-1}{{z}}^{-n}, |{{z}}|<1.
\label{deltaDpm_sq}
\end{eqn}

\subsubsection{(a),(b)}
For the case of (a) ${\mathfrak{S}\hspace{-.4ex}}{\mathbin{\substack{{\bullet}\\{\circ}{\bullet}}}}$ and (b) ${\mathfrak{S}\hspace{-.4ex}}{\mathbin{\substack{\circ\\\circ\bullet}}}$ of Fig. \ref{stepboundarywaveguides_sq}, it follows from the expression for the incident wave mode \eqref{uinc_sq}, the Wiener--Hopf equations for the right incidence \eqref{WHKeq_sq_X} and the left incidence \eqref{WHKeq_sq_X_altinc}, as well as the combined notation for the right hand side \eqref{Cs_sq_def} and the multiplicative factorization ${{\mathpzc{L}}}={{\mathpzc{L}}}_{+}{{\mathpzc{L}}}_{-}$, 
that ${{\mathpzc{L}}}_{+}({{z}}){\su}_{1;+}({{z}})+{{\mathpzc{L}}}^{-1}_{-}({{z}}){\su}_{1;-}({{z}})={{\mathpzc{C}}}({{z}}), 
 {{z}}\in{{\mathscr{A}}}$ where
\beqan
{{\mathpzc{C}}}({{z}})&=&({{\mathpzc{L}}}_{+}({{z}})-{{\mathpzc{L}}}^{-1}_{-}({{z}})){{\mathrm{A}}}{{a}}_{({{{\kappa}}^{{\text{in}}}}){1}}(\delta_{D-}({{z}} {{z}}_{{\text{in}}}^{-1})\delta_{s,{A}}
-\delta_{D+}({{z}}{{z}}_{{\text{in}}}^{-1})\delta_{s,{B}}), 
\quad {{z}}\in{{\mathscr{A}}},
\label{Cz_ab}\\
\text{where }
{{z}}_{{\text{in}}}&{:=}& e^{-i{\upkappa}_x}\in\mathbb{C}.
\label{zPdef_sq}
\eeqan
\begin{remark}
By virtue of the chosen direction of incidence from the right portion ${{\mathfrak{s}}}={A},$ and the assumption of dissipation (${\upomega}_2>0$), it is concluded that $|{{z}}_{{\text{in}}}|>1$ when $s={A}$ against the case when $s={B}$ and the chosen direction of incidence from the left portion ${{\mathfrak{s}}}={B},$ so that $|{{z}}_{{\text{in}}}|<1$.
\label{rem_zP} 
\end{remark}
In terms of the one-sided discrete Fourier transform, 
\begin{eqn}
{\su}_{1;\pm}({{z}})={{\mathpzc{C}}}_{\pm}({{z}}){{\mathpzc{L}}}_{\pm}^{\mp1}({{z}}), {{z}}\in\mathbb{C}, |{{z}}|\gtrless
{{\mathit{R}}}_{\pm},
\label{u1pmsol_ab}
\end{eqn}
\begin{eqn}
\text{and after simplifying further, }
{\su}_{1}^F({{z}})&={{\mathrm{A}}}{\mathtt{C}}_0\frac{{{z}}{{\mathpzc{K}}}({{z}})}{{{z}}-{{z}}_{{\text{in}}}}, 
{{\mathpzc{K}}}({{z}}){:=}\frac{{{\mathpzc{L}}}({{z}})-1}{{{\mathpzc{L}}}_{+}({{z}})}, \\&
{\mathtt{C}}_0{:=}-{{a}}_{({{{\kappa}}^{{\text{in}}}}){1}}({{\mathpzc{L}}}_{+}({{z}}_{{\text{in}}})\delta_{s,{A}}+{{\mathpzc{L}}}^{-1}_{-}({{z}}_{{\text{in}}})\delta_{s,{B}})\in\mathbb{C},
\label{u1zsol_ab}
\end{eqn}
for ${{z}}\in{{\mathscr{A}}}$. 

\subsubsection{(c),(d)}
In the case of ${\mathfrak{S}\hspace{-.4ex}}{\mathbin{\substack{\circ\\\bullet\circ}}}$ and ${\mathfrak{S}\hspace{-.4ex}}{\mathbin{\substack{\bullet\\\bullet\circ}}}$, it follows from \eqref{uinc_sq}, \eqref{WHCeq_sq_R}, \eqref{WHCeq_sq_R_altinc}, and \eqref{Cs_sq_def}, using ${{\mathpzc{L}}}={{\mathpzc{L}}}_{+}{{\mathpzc{L}}}_{-}$, 
that ${{\mathpzc{L}}}_{+}{\su}_{2;+}+{{\mathpzc{L}}}_{-}^{-1}{{\su}_{2;-}}={{\mathpzc{C}}}({{z}}), 
 {{z}}\in{{\mathscr{A}}}$ where the right hand side is
\begin{eqn}
{{\mathpzc{C}}}({{z}})
&=({{\mathpzc{L}}}_{-}^{-1}-{{\mathpzc{L}}}_{+})({{\mathpzc{W}}}-({{\mathpzc{Q}}}-1){{\mathrm{A}}}{{a}}_{({{{\kappa}}^{{\text{in}}}}){1}}\delta_{D-}({{z}}{{z}}_{{\text{in}}}^{-1})\delta_{s,{A}}\\&+{{\mathrm{A}}}{{a}}_{({{{\kappa}}^{{\text{in}}}}){2}}\delta_{D+}({{z}}{{z}}_{{\text{in}}}^{-1})\delta_{s,{B}}),
\quad {{z}}\in{{\mathscr{A}}}, \label{Cz_cd}
\end{eqn}
where ${\mathpzc{W}}$ is defined by \eqref{q2beta_R}, i.e. ${\su}_{-1, 1}-{{z}} {\su}_{0, 1}$.
In terms of the one-sided discrete Fourier transform the complex function ${\su}_{2}^F$ is given by
\begin{eqn}
{\su}_{2;\pm}({{z}})={{\mathpzc{C}}}_{\pm}({{z}}){{\mathpzc{L}}}_{\pm}^{\mp1}({{z}}), {{z}}\in\mathbb{C}, |{{z}}|\gtrless
{{\mathit{R}}}_{\pm},
\label{u2pmsol_cd}
\end{eqn}
\begin{eqn}
\text{and, similar to \eqref{u1zsol_ab}},
{\su}_{1}^F({{z}})&={{\mathrm{A}}}{\mathtt{C}}_0\frac{{{z}}{{\mathpzc{K}}}({{z}})}{{{z}}-{{z}}_{{\text{in}}}}, 
{{\mathpzc{K}}}({{z}}){:=}\frac{1}{(1-{z}_{\sqh}{z}^{-1}){{\mathpzc{L}}}_{+}({{z}})}\delta_{s,{A}}, \\
{\mathtt{C}}_0&{:=}{{a}}_{({{{\kappa}}^{{\text{in}}}}){1}}\frac{({\mathpzc{Q}}({z}_{{\text{in}}})-1){{\mathpzc{L}}}_{+}({{z}}_{{\text{in}}})}{{z}_{\sqh}^{-1}-{{z}}_{{\text{in}}}}\delta_{s,{A}}
+{{a}}_{({{{\kappa}}^{{\text{in}}}}){2}}\frac{{{\mathpzc{L}}}_{-}^{-1}({{z}}_{{\text{in}}})}{{{z}}^{-1}_{\sqh}-{{z}}_{{\text{in}}}}\delta_{s,{B}}\in\mathbb{C},
\label{u2zsol_cd}
\end{eqn}
for ${{z}}\in{{\mathscr{A}}}$. Note that ${{z}}_{\sqh}$ and ${{z}}_{\sqh}^{-1}$ are the two zeros of ${\mathpzc{Q}}-1$ (with $|{{z}}_{\sqh}|<1$), i.e. ${\mathpzc{Q}}({z})-1={z}_{\sqh}^{-1}(1-{z}_{\sqh}{z})(1-{z}_{\sqh}{z}^{-1})=-{z}^{-1}({z}-{z}_{\sqh})({z}-{z}_{\sqh}^{-1})$; $({\mathpzc{Q}}-1)_\pm({z})={z}_{\sqh}^{-1/2}(1-{z}_{\sqh}{z}^{\mp1})$.

\subsubsection{(e),(f)}
In the case of ${\mathfrak{S}\hspace{-.4ex}}{\mathbin{\substack{\circ\\\circ\circ}}}$ and ${\mathfrak{S}\hspace{-.4ex}}{\mathbin{\substack{\bullet\\\circ\circ}}}$, it follows from \eqref{uinc_sq}, \eqref{WHKeq_sq_R}, \eqref{WHKeq_sq_R_altinc}, and \eqref{Cs_sq_def}, after applying the multiplicative factorization ${{\mathpzc{L}}}={{\mathpzc{L}}}_{+}{{\mathpzc{L}}}_{-}$, 
that ${{\mathpzc{L}}}_{+}{\su}_{2;+}+{{\mathpzc{L}}}_{-}^{-1}{\su}_{2;-}={{\mathpzc{C}}}({{z}}), 
 {{z}}\in{{\mathscr{A}}}$ where
 \begin{eqn}
{{\mathpzc{C}}}({{z}})
&= ({{\mathpzc{L}}}_{+}({{z}})-{{\mathpzc{L}}}^{-1}_{-}({{z}})){{\mathrm{A}}}{{a}}_{({{{\kappa}}^{{\text{in}}}}){2}}(\delta_{D-}({{z}}{{z}}_{{\text{in}}}^{-1})\delta_{s,{A}}
-\delta_{D+}({{z}}{{z}}_{{\text{in}}}^{-1})\delta_{s,{B}})\\
&+({{\mathpzc{L}}}_{+}({{z}})-{{\mathpzc{L}}}^{-1}_{-}({{z}})){{\mathpzc{H}}}^{-1}((1-{z}){\su}^{{\text{in}}}_{0, 1}\delta_{s, {A}}
+(1-{z}){\su}_{0, 1}), 
\quad {{z}}\in{{\mathscr{A}}}.
\end{eqn}
In terms of the one-sided discrete Fourier transform the complex function ${\su}_{1}^F$ is given by
\begin{eqn}
{\su}_{2;\pm}({{z}})={{\mathpzc{C}}}_{\pm}({{z}}){{\mathpzc{L}}}_{\pm}^{\mp1}({{z}}), {{z}}\in\mathbb{C}, |{{z}}|\gtrless{{\mathit{R}}}_{\pm}.
\label{u2pmsol_ef}
\end{eqn}
\begin{eqn}
\text{Thus, }
{\su}_{2}^F({{z}})
&={{\mathrm{A}}}{\mathtt{C}}_0{{{z}}{{\mathpzc{K}}}({{z}})}(\frac{1}{{{z}}-{{z}}_{{\text{in}}}}-{{z}}^{-1}\frac{1}{1-{z}_{\sqh}{{z}}_{{\text{in}}}}\frac{\mathcal{F}({z})}{\mathcal{F}({z}_{\sqh}^{-1})}), {{\mathpzc{K}}}({{z}}){:=}\frac{{{\mathpzc{L}}}({{z}})-1}{{{\mathpzc{L}}}_{+}({{z}})}, \\
\mathcal{F}({z})&{:=}\big((1-{z})\big(\frac{{{{\mathpzc{L}}}}^{-1}_-({z}_{{\mathpzc{h}}})}{{\mathpzc{H}}_-({z}_{{\mathpzc{h}}}){\mathpzc{H}}_+({{z}})}+\frac{{{{\mathpzc{L}}}}_+({z}_{{\mathpzc{h}}}^{-1})}{{\mathpzc{H}}_+({z}_{{\mathpzc{h}}}^{-1}){\mathpzc{H}}_-({{z}})}\big)-{\mathpzc{C}}^a_-(0)-{z}{\mathpzc{C}}^a_+(\infty)\big)\\
{\mathtt{C}}_0&{:=}-{{a}}_{({{{\kappa}}^{{\text{in}}}}){2}}({{\mathpzc{L}}}_{+}({{z}}_{{\text{in}}})\delta_{s,{A}}+{{\mathpzc{L}}}^{-1}_{-}({{z}}_{{\text{in}}})\delta_{s,{B}})\in\mathbb{C}.
\label{u2zsol_ef_solution}
\end{eqn}

\subsubsection{(g),(h)}
In the case of ${\mathfrak{S}\hspace{-.4ex}}{\mathbin{\substack{\bullet\\\bullet\bullet}}}$ and ${\mathfrak{S}\hspace{-.4ex}}{\mathbin{\substack{\circ\\\bullet\bullet}}}$, by inspection it is clear that the relevant equations (after minor changes) coincide with the bifurcated waveguides ${\mathfrak{S}\hspace{-.4ex}}{\mathbin{\substack{\bullet}{\hspace{-.2ex}}\substack{\bullet\\ \bullet}}}$ 
and ${\mathfrak{S}\hspace{-.4ex}}{\mathbin{\substack{\bullet}{\hspace{-.2ex}}\substack{\bullet\\ \circ}}}$ (with ${\mathtt{N}_{\mathfrak{a}}}=0, {\mathtt{N}_{\mathfrak{b}}}={\mathtt{N}}-1$) analyzed by \cite{Bls9s} for the incidence from the broader side of the step, i.e., ${{\mathfrak{s}}}={A}$. 
It follows from \eqref{uinc_sq}, \eqref{WHCeq_sq_X}, \eqref{WHCeq_sq_X_altinc}, and \eqref{Cs_sq_def}, after applying the multiplicative factorization ${{\mathpzc{L}}}={{\mathpzc{L}}}_{+}{{\mathpzc{L}}}_{-}$, 
that ${{\mathpzc{L}}}_{+}{\su}_{2;+}+{{{\mathpzc{L}}}_{-}^{-1}}{{\su}_{2;-}}={{\mathpzc{C}}}({{z}}), 
 {{z}}\in{{\mathscr{A}}}$ where the right hand side is
\begin{eqn}
{{\mathpzc{C}}}({{z}})
&=({{\mathpzc{L}}}_{-}^{-1}-{{\mathpzc{L}}}_{+})({{\mathpzc{W}}}-{{\mathpzc{Q}}}{{a}}_{({{{\kappa}}^{{\text{in}}}}){1}}\delta_{D-}({{z}}{{z}}_{{\text{in}}}^{-1})\delta_{s,{A}}
+{{a}}_{({{{\kappa}}^{{\text{in}}}}){2}}\delta_{D+}({{z}}{{z}}_{{\text{in}}}^{-1})\delta_{s,{B}}),\quad {{z}}\in{{\mathscr{A}}}. \label{Cz_gh}
\end{eqn}
and ${\mathpzc{W}}$ is defined by \eqref{q2beta_X}. Recall Remark \ref{rem_zP}. 

In terms of the one-sided discrete Fourier transform the complex function ${\su}_{2}^F$ is given by
\begin{eqn}
{\su}_{2;\pm}({{z}})={{\mathpzc{C}}}_{\pm}({{z}}){{\mathpzc{L}}}_{\pm}^{\mp1}({{z}}), {{z}}\in\mathbb{C}, |{{z}}|\gtrless{{\mathit{R}}}_{\pm},
\label{u2pmsol_gh}
\end{eqn}
\begin{eqn}
\text{while, similar to \eqref{u1zsol_ab}},
{\su}_{1}^F({{z}})&={{\mathrm{A}}}{\mathtt{C}}_0\frac{{{z}}{{\mathpzc{K}}}({{z}})}{{{z}}-{{z}}_{{\text{in}}}}, 
{{\mathpzc{K}}}({{z}}){:=}\frac{1}{(1-{z}_{\sq}{z}^{-1}){{\mathpzc{L}}}_{+}({{z}})}\delta_{s,{A}}, \\
{\mathtt{C}}_0&{:=}{{a}}_{({{{\kappa}}^{{\text{in}}}}){1}}\frac{{\mathpzc{Q}}({z}_{{\text{in}}}){{\mathpzc{L}}}_{+}({{z}}_{{\text{in}}})}{{z}_{\sq}^{-1}-{{z}}_{{\text{in}}}}\delta_{s,{A}}
+{{a}}_{({{{\kappa}}^{{\text{in}}}}){2}}\frac{{{\mathpzc{L}}}_{-}^{-1}({{z}}_{{\text{in}}})}{{{z}}^{-1}_{\sq}-{{z}}_{{\text{in}}}}\delta_{s,{B}}.
\label{u1zsol_gh}
\end{eqn}

\subsubsection{(i),(j)}
Let 
${{\mathpzc{P}}}{:=}{\mathpzc{Q}}-1={{\mathpzc{H}}}+1.$
In the case of ${\mathfrak{S}\hspace{-.4ex}}{\mathbin{\substack{\circ\\\circ\overline{\bullet}}}}$ and ${\mathfrak{S}\hspace{-.4ex}}{\mathbin{\substack{{\bullet}\\{\circ}\overline{\bullet}}}}$, it follows from \eqref{uinc_sq}, \eqref{WHKeq_sq_RX2}, \eqref{WHKeq_sq_RX2_altinc}, and \eqref{Cs_sq_def}, after applying ${{\mathpzc{L}}}={{\mathpzc{L}}}_{+}{{\mathpzc{L}}}_{-}$, 
that ${{\mathpzc{L}}}_{+}{\su}_{2;+}+{{\mathpzc{L}}}_{-}^{-1}{\su}_{2;-}={{\mathpzc{C}}}({{z}}), 
 {{z}}\in{{\mathscr{A}}}$ where
 \begin{eqn}
{{\mathpzc{C}}}({{z}})
&= ({{\mathpzc{L}}}_{+}({{z}})-{{\mathpzc{L}}}^{-1}_{-}({{z}})){{\mathrm{A}}}{{a}}_{({{{\kappa}}^{{\text{in}}}}){2}}(\delta_{D-}({{z}}{{z}}_{{\text{in}}}^{-1})\delta_{s,{A}}-\delta_{D+}({{z}}{{z}}_{{\text{in}}}^{-1})\delta_{s,{B}})\\
&+({{\mathpzc{L}}}_{+}({{z}})-{{\mathpzc{L}}}^{-1}_{-}({{z}})){{{\mathpzc{P}}}}^{-1}((1-{z}){\su}^{{\text{in}}}_{0, 1}\delta_{s, {A}}+(1-{z}){\su}_{0, 1}), 
\quad {{z}}\in{{\mathscr{A}}}.
\end{eqn}
In terms of the one-sided discrete Fourier transform the complex function ${\su}_{2}^F$ is given by
\begin{eqn}
{\su}_{2;\pm}({{z}})={{\mathpzc{C}}}_{\pm}({{z}}){{\mathpzc{L}}}_{\pm}^{\mp1}({{z}}), {{z}}\in\mathbb{C}, |{{z}}|\gtrless{{\mathit{R}}}_{\pm},
\label{u2pmsol_ef2}
\end{eqn}
\begin{eqn}
\text{while }{\su}_{2}^F({{z}})&={{\mathrm{A}}}{\mathtt{C}}_0{{{z}}{{\mathpzc{K}}}({{z}})}(\frac{1}{{{z}}-{{z}}_{{\text{in}}}}-{{z}}^{-1}\frac{1}{1-{z}_{\sq}{{z}}_{{\text{in}}}}\frac{\mathcal{F}({z})}{\mathcal{F}({z}_{\sq}^{-1})}), {{\mathpzc{K}}}({{z}}){:=}\frac{{{\mathpzc{L}}}({{z}})-1}{{{\mathpzc{L}}}_{+}({{z}})}, \\
\mathcal{F}({{z}})&=\big((1-{z})\big(\frac{{{\mathpzc{L}}}^{-1}_-({z}_{{\sqh}})}{{{\mathpzc{P}}}_-({z}_{{\sqh}}){{\mathpzc{P}}}_+({{z}})}+\frac{{{\mathpzc{L}}}_+({z}_{{\sqh}}^{-1})}{{{\mathpzc{P}}}_+({z}_{{\sqh}}^{-1}){{\mathpzc{P}}}_-({{z}})}\big)-{\mathpzc{C}}^a_-(0)-{z}{\mathpzc{C}}^a_+(\infty)\big), \\
{\mathtt{C}}_0&{:=}-{{a}}_{({{{\kappa}}^{{\text{in}}}}){2}}({{\mathpzc{L}}}_{+}({{z}}_{{\text{in}}})\delta_{s,{A}}+{{\mathpzc{L}}}^{-1}_{-}({{z}}_{{\text{in}}})\delta_{s,{B}})\in\mathbb{C}.
\label{u2zsol_ef2_solution}
\end{eqn}

Combining \eqref{u1pmsol_ab} with the expression \eqref{bulk_k_sq_X} for ${\mathfrak{S}\hspace{-.4ex}}{\mathbin{\substack{{\bullet}\\{\circ}{\bullet}}}}$ in \S\ref{crack_sq_X} (and \eqref{bulk_k_sq_RX} for ${\mathfrak{S}\hspace{-.4ex}}{\mathbin{\substack{\circ\\\circ\bullet}}}$ in \S\ref{crack_sq_RX}), \eqref{u2pmsol_cd} with \eqref{bulk_k_sq_RX}, \eqref{u2Fu1F_sq_RX} for ${\mathfrak{S}\hspace{-.4ex}}{\mathbin{\substack{\circ\\\bullet\circ}}}$ in \S\ref{slit_sq_R} (and \eqref{bulk_k_sq_RX}, \eqref{u2Fu1F_sq_X} for ${\mathfrak{S}\hspace{-.4ex}}{\mathbin{\substack{\bullet\\\bullet\circ}}}$ in \S\ref{slit_sq_XR}), \eqref{u2pmsol_ef} with \eqref{bulk_k_sq_RX}, \eqref{u3Fu2F_sq_R} for ${\mathfrak{S}\hspace{-.4ex}}{\mathbin{\substack{\circ\\\circ\circ}}}$ in \S\ref{crack_sq_R} (and \eqref{bulk_k_sq_RX}, \eqref{u2Fu1F_sq_RX} for ${\mathfrak{S}\hspace{-.4ex}}{\mathbin{\substack{\bullet\\\circ\circ}}}$ in \S\ref{crack_sq_XR}), \eqref{u2pmsol_gh} with \eqref{bulk_k_sq_X}, \eqref{u2Fu1F_sq_X} for ${\mathfrak{S}\hspace{-.4ex}}{\mathbin{\substack{\bullet\\\bullet\bullet}}}$ in \S\ref{slit_sq_X} (and \eqref{bulk_k_sq_RX}, \eqref{u2Fu1F_sq_RX} for ${\mathfrak{S}\hspace{-.4ex}}{\mathbin{\substack{\circ\\\bullet\bullet}}}$ in \S\ref{slit_sq_RX}), ${\su}_{{\mathtt{x}}, {\mathtt{y}}}$ is determined by inverse Fourier transform \cite{jury,Slepyanbook}, 
\begin{eqn}
{\su}_{{\mathtt{x}}, {\mathtt{y}}}&=\frac{1}{2\pi i}\oint_{{{\mathcal{C}}}_{{z}}}{\su}_{{\mathtt{y}}}^{F}({{z}})d{{z}},
\label{umnsolk}
\end{eqn}
where ${{\mathcal{C}}}_{{z}}$ is a rectifiable, closed, counterclockwise contour in the annulus ${{\mathscr{A}}}$. The integration contour ${{\mathcal{C}}}_{{z}}$ in \eqref{umnsolk} may be chosen as the unit circle $\mathbb{T}\subset\mathbb{C}$ when ${\upomega}_2>0$ \cite{Bls9s}. These expressions can be simplified, though the details are passed over in order to maintain the focus on the waveguide transmission property rather than the exact solution in the waveguide. 
Employing \eqref{umnsolk}, above description provides the complete solution of the wave propagation problem in integral form (few extra details of Wiener--Hopf method have been also provided in the supplementary \S2 and graphical comparison between exact solution at one site with the numerical solution is provided in the supplementary \S3). 

\section{Reflectance and transmittance}
\label{reftrans}
At distances large compared to the width of the waveguides and away from the step, the far-field approximation of the exact solution \eqref{umnsolk} is used to calculate the physically interesting entities associated with the fraction of incident energy flux \cite{Brillouin} which is transmitted across the atomic step discontinuity.
This is accomplished by summing the energy in outgoing wavemodes ahead and behind the step. In fact, the reflectance (resp. transmittance) is defined as the ratio of the total energy flux in outgoing wavemodes ahead of (resp. behind) the step vs the energy flux carried for a specific incident wave mode \cite{Brillouin}. 
Using the dispersion relation ${\upomega}^2=4(\sin^2{\frac{1}{2}}{\upxi}+\sin^2{\frac{1}{2}}{\upeta}), {\upxi}\in [-\pi, \pi],$ in the square lattice waveguides \cite{Bls9}, since ${\upeta}$ is constant for a wave mode, it is easy to see that
${\mathit{v}}({\upxi}){:=}\pd{{\upomega}}{{\upxi}}={\upomega}^{-1}\sin{\upxi}$, where ${\mathit{v}}$ denotes the group velocity \cite{Brillouin} of the propagating wave with wave number ${\upxi}$. The energy flux in a wave mode, as a product of energy density and the group velocity, can be easily calculated (see \cite{Brillouin}); for instance for the incident wave mode 
it is given by
\begin{eqn}
{\mathscr{E}^{\text{in}}}&=\sum_{{\mathtt{y}}\in\mathbb{Z}_1^{{\mathtt{N}}}}|{{\mathrm{A}}}|^2|{{a}}_{({{\kappa}}^{{\text{in}}}){{\mathtt{y}}}}|^2{\mathit{v}}({\upxi}_{{\text{in}}})=|{{\mathrm{A}}}|^2{\mathit{v}}({\upxi}_{{\text{in}}})={\upomega}^{-1}|{{\mathrm{A}}}|^2\sin{\upxi}_{{\text{in}}}.
\label{energyflux_inc}
\end{eqn}
The sets of ${z}_{{\ast}}$ corresponding to the outgoing waves ahead and behind the step, respectively, are
\begin{eqn}
{{\mathcal{Z}}}^+=\{{z}\in\mathbb{T} \big| {{\mathscr{N}}_+({{z}})=0}\}, \quad{{\mathcal{Z}}}^{-}=\{{z}\in\mathbb{T} \big| {{\mathscr{D}}_-({{z}})=0}\},
\label{Zer_sq_c}
\end{eqn}
while those 
corresponding to incoming waves are
$\widetilde{{\mathcal{Z}}}^-=\{{z}\in\mathbb{T} \big| {{\mathscr{N}}_-({{z}})=0}\}$ for ${\mathfrak{s}}={A}$ and $\widetilde{{\mathcal{Z}}}^+=\{{z}\in\mathbb{T} \big| {{\mathscr{D}}_+({{z}})=0}\}$ for ${\mathfrak{s}}={B}$. 
Note that $\#{{\mathcal{Z}}}^\pm=\#{{\mathcal{Z}}}^\pm$, etc.

\subsection{Case (a)}
Consider the waveguide (a), i.e., ${\mathfrak{S}\hspace{-.4ex}}{\mathbin{\substack{{\bullet}\\{\circ}{\bullet}}}}.$ 
Since, by \eqref{umnsolk},
$\su_{{\mathtt{x}}, 1}=\frac{1}{2\pi i}\oint_{\mathbb{T}} \su_{1}^F({{z}}){{z}}^{{\mathtt{x}}-1}d{{z}}, {\mathtt{x}}\in\mathbb{Z},$
in particular
where $\su_{1}^F=\su_{1;+}+\su_{1;-}$, as ${\mathtt{x}}\to\pm\infty$, the asymptotic expression for $\su_{1}$ can be obtained by simply analyzing 
$\su_{{\mathtt{x}},1}=\frac{1}{2\pi i}\oint_{\mathbb{T}} \su_{1;\pm}({{z}}){{z}}^{{\mathtt{x}}-1}d{{z}}, {\mathtt{x}}\in\mathbb{Z}^{\pm}.$
Indeed, after deforming contour of integration and applying residue calculus, using \eqref{u1pmsol_ab}, 
an exact expression is given by 
(assuming ${\upomega}_2\to0+$)
\begin{eqn}
\su_{{\mathtt{x}},1}=\pm\sum\nolimits_{{{z}}={{z}}_{{\ast}}, |{{z}}_{{\ast}}|\lessgtr1} \text{Res }{{\mathpzc{C}}}_{\pm}({{z}}){{\mathpzc{L}}}_{\pm}^{\mp1}({{z}}){{z}}^{{\mathtt{x}}-1}, \quad\quad {\mathtt{x}}\in\mathbb{Z}^{\pm}.
\label{um1asym1}
\end{eqn}
The asymptotic approximation of $\su_{{\mathtt{x}},1}$ takes into account only the contributions due to ${z}_{{\ast}}$ corresponding to the outgoing wave modes (including a reflected wave), as expected. 

Thus, for ${\mathfrak{s}}={A},$
\begin{eqn}
\su_{{\mathtt{x}},1}&\sim{{\mathrm{A}}}{{a}}_{({{{\kappa}}^{{\text{in}}}}){1}}\frac{{\mathscr{N}}_+({{z}}_{{\text{in}}})}{{\mathscr{D}}_+({{z}}_{{\text{in}}})}\sum\nolimits_{{{{z}}_{{\ast}} \in{{\mathcal{Z}}}^+}}\frac{{\mathscr{D}}_+({{z}}_{{\ast}})}{{\mathscr{N}}'_+({{z}}_{{\ast}})}\frac{{{z}}_{{\ast}}^{{\mathtt{x}}}}{{{z}}_{{\ast}}-{{z}}_{{\text{in}}}}, {\mathtt{x}}\to+\infty,\\
\su_{{\mathtt{x}},1}&\sim{{\mathrm{A}}}{{a}}_{({{{\kappa}}^{{\text{in}}}}){1}}(-{{z}}_{{\text{in}}}^{{\mathtt{x}}}+\frac{{\mathscr{N}}_+({{z}}_{{\text{in}}})}{{\mathscr{D}}_+({{z}}_{{\text{in}}})}\sum\nolimits_{{{{z}}_{{\ast}} \in{{\mathcal{Z}}}^-}}\frac{{\mathscr{N}}_-({{z}}_{{\ast}})}{{\mathscr{D}}'_-({{z}}_{{\ast}})}\frac{{{z}}_{{\ast}}^{{\mathtt{x}}}}{{{z}}_{{\ast}}-{{z}}_{{\text{in}}}}), {\mathtt{x}}\to-\infty.
\label{vm0asym}
\end{eqn}
Note that in the expression for ${\mathtt{x}}\to+\infty$, ${{z}}_{{\text{in}}}^{-1}$ is included in the sum (here ${\mathscr{N}}_+({{z}}_{{\text{in}}}^{-1})=0, {\mathscr{N}}_-({{z}}_{{\text{in}}}^{-1})\ne0$ but ${\mathscr{N}}_-({{z}}_{{\text{in}}})=0$). 
In the expression for ${\mathtt{x}}\to-\infty$, ${{z}}_{{\text{in}}}$ does not occur in the sum (${\mathscr{N}}_-({{z}}_{{\text{in}}})=0, {\mathscr{N}}_+({{z}}_{{\text{in}}})\ne0$ but ${\mathscr{N}}_+({{z}}_{{\text{in}}}^{-1})=0$). 
Using \eqref{vm0asym}, after simplification, the total field is given by $\su^{{t}}_{{\mathtt{x}},1}=\su^{{\text{in}}}_{{\mathtt{x}},1}+\su_{{\mathtt{x}},1}$. 
Thus, for a wave incident from the right side, the transmitted waves includes the contribution from the residue associated with the incident wave (which cancels another equal contribution), while the reflected waves includes the contribution from the residue associated with the reflected wave. 

Using an expansion of the wave field in terms of the outgoing wave modes, it is possible to construct 
the far-field in terms of the eigenmodes associated with the two different portions of the lattice strip. The eigenmodes for a square lattice strip with fixed or free boundary are well known (see also \cite{Bls9} for a systematic catalogue). 
Thus, it is required that ahead and behind the step (with ${{\kappa}}({{z}})$ denoting the index of wave mode corresponding to ${z}$)
\begin{eqn}
\su_{{\mathtt{x}}, {\mathtt{y}}}&\sim{{\mathrm{A}}}\sum\nolimits_{{{{z}}\in{{\mathcal{Z}}}^+}}{{\mathrm{A}}}_{{{\kappa}}({{z}})}{{a}}_{+({{\kappa}}({{z}})){{\mathtt{y}}}}{{z}}^{{\mathtt{x}}}, {\mathtt{x}}\to+\infty,\\
\su_{{\mathtt{x}}, {\mathtt{y}}}&\sim{{\mathrm{A}}}\sum\nolimits_{{{z}}\in{{\mathcal{Z}}}^-}{{\mathrm{A}}}_{{{\kappa}}({{z}})}{{a}}_{-({{\kappa}}({{z}})){{\mathtt{y}}}}{{z}}^{{\mathtt{x}}}-{{\mathrm{A}}}{{a}}_{({{{\kappa}}^{{\text{in}}}}){{\mathtt{y}}}}{{z}}_{{\text{in}}}^{{\mathtt{x}}}, {\mathtt{x}}\to-\infty.
\label{modalexpansion}
\end{eqn}
Comparing \eqref{vm0asym} with the expression of $\su_{\cdot,1}$, the coefficients ${{\mathrm{A}}}_{{{\kappa}}({{z}})}$ \eqref{modalexpansion} can be found:
\begin{eqn}{{\mathrm{A}}}_{{{\kappa}}({{z}})}&=\frac{{{a}}_{({{{\kappa}}^{{\text{in}}}}){1}}}{{{a}}_{+({{\kappa}}({{z}})){1}}}\frac{1}{{{z}}-{{z}}_{{\text{in}}}} \frac{{\mathscr{D}}_+({{z}}){\mathscr{N}}_+({{z}}_{{\text{in}}})}{{\mathscr{N}}'_+({{z}}){\mathscr{D}}_+({{z}}_{{\text{in}}})},\\{{\mathrm{A}}}_{{{\kappa}}({{z}})}&=\frac{{{a}}_{({{{\kappa}}^{{\text{in}}}}){1}}}{{{a}}_{-({{\kappa}}({{z}}))1}}\frac{1}{{{z}}-{{z}}_{{\text{in}}}} \frac{{\mathscr{N}}_-({{z}}){\mathscr{N}}_+({{z}}_{{\text{in}}})}{{\mathscr{D}}'_-({{z}}){\mathscr{D}}_+({{z}}_{{\text{in}}})}.\end{eqn}
Using above equation, hence, it follows that
the total displacement field is given by
\begin{eqn}
\su^{{t}}_{{\mathtt{x}}, {\mathtt{y}}}&\sim{{\mathrm{A}}}{{a}}_{({{{\kappa}}^{{\text{in}}}}){{\mathtt{y}}}}{{z}}_{{\text{in}}}^{{\mathtt{x}}}+{{\mathrm{A}}}\frac{{\mathscr{N}}_+({{z}}_{{\text{in}}})}{{\mathscr{D}}_+({{z}}_{{\text{in}}})}\sum\nolimits_{{{z}}\in{{\mathcal{Z}}}^+}
\frac{{{a}}_{({{{\kappa}}^{{\text{in}}}}){1}}{{a}}_{+({{\kappa}}({{z}})){{\mathtt{y}}}}{{z}}^{{\mathtt{x}}}}{{{a}}_{+({{\kappa}}({{z}}))1}}\frac{1}{{{z}}-{{z}}_{{\text{in}}}} \frac{{\mathscr{D}}_+({{z}})}{{\mathscr{N}}'_+({{z}})}\\
\su^{{t}}_{{\mathtt{x}}, {\mathtt{y}}}&\sim{{\mathrm{A}}}\frac{{\mathscr{N}}_+({{z}}_{{\text{in}}})}{{\mathscr{D}}_+({{z}}_{{\text{in}}})}\sum\nolimits_{{{z}}\in{{\mathcal{Z}}}^-}
\frac{{{a}}_{({{{\kappa}}^{{\text{in}}}}){1}}{{a}}_{-({{\kappa}}({{z}})){{\mathtt{y}}}}{{z}}^{{\mathtt{x}}}}{{{a}}_{-({{\kappa}}({{z}}))1}}\frac{1}{{{z}}-{{z}}_{{\text{in}}}} \frac{{\mathscr{N}}_-({{z}})}{{\mathscr{D}}'_-({{z}})},
\label{farfield_k_sq}
\end{eqn}
as ${\mathtt{x}}\to+\infty$ and ${\mathtt{x}}\to-\infty$, respectively. 
Since the part of the strip ahead the step is fixed (of width ${\mathtt{N}}$), i.e. ${\mathfrak{S}{\sbfrac{\bullet}{\bullet}}}$, it is easy to see that
${{a}}_{+({{\kappa}}){1}}=\sqrt{{2}/{{{\mathtt{N}}}+1}}\sin{{\upeta}k},$
${{{a}}_{({{{\kappa}}^{{\text{in}}}}){1}}}={{a}}_{+({\kappa}_{{{z}}_{{\text{in}}}}){1}},$
and for the fixed-free strip of type ${\mathfrak{S}{\sbfrac{\bullet}{\circ}}}$ behind the step, ${{a}}_{-({{\kappa}}){1}}=({2}/{\sqrt{2{\mathtt{N}}+1}})\cos{{\frac{1}{2}}}{{\upeta}k}$.
The energy flux in each outgoing wave mode is straighforward. as the expression is same as that stated in \eqref{energyflux_inc} \cite{Brillouin}.
Using the fact that only linear waves are considered, the reflectance and the transmittance of bifurcated wave guide (say, for a wave incident from the right side) are given by
\begin{eqn}
{\mathscr{R}}({{z}}_{{\text{in}}})={\mathscr{E}^{\text{in}}}^{-1}\sum\nolimits_{{{z}}\in{{\mathcal{Z}}}^+}{\mathscr{E}}({{\kappa}}({{z}}))&=\frac{(2\sin{\upxi}_{{\text{in}}})^{-1}}{|{{\mathpzc{L}}}^{-1}_{+}({{z}}_{{\text{in}}})|^2}\sum\nolimits_{{{z}}\in{{\mathcal{Z}}}^+}\bigg|\frac{{{a}}_{({{{\kappa}}^{{\text{in}}}}){1}}}{{{a}}_{+({{\kappa}}({{z}}))1}}\bigg|^2\frac{(-2\sin{\upxi})}{|{{z}}-{{z}}_{{\text{in}}}|^2} \bigg|\frac{{\mathscr{D}}_+({{z}})}{{\mathscr{N}}'_+({{z}})}\bigg|^2,\\
{\mathscr{T}}({{z}}_{{\text{in}}})={\mathscr{E}^{\text{in}}}^{-1}\sum\nolimits_{{{z}}\in{{\mathcal{Z}}}^-}{\mathscr{E}}({{\kappa}}({{z}}))&=\frac{(2\sin{\upxi}_{{\text{in}}})^{-1}}{|{{\mathpzc{L}}}^{-1}_{+}({{z}}_{{\text{in}}})|^2}\sum\nolimits_{{{z}}\in{{\mathcal{Z}}}^-}\bigg|\frac{{{a}}_{({{{\kappa}}^{{\text{in}}}}){1}}}{{{a}}_{-({{\kappa}}({{z}}))1}}\bigg|^2\frac{(2\sin{\upxi})}{|{{z}}-{{z}}_{{\text{in}}}|^2} \bigg|\frac{{\mathscr{N}}_-({{z}})}{{\mathscr{D}}'_-({{z}})}\bigg|^2.
\label{RTk_sq_X}
\end{eqn}

By repeated application of identities in Appendix C of \cite{Bls5k_tube} (or Appendix D of \cite{Bls9s}), above expressions can be simplified further. It noteworthy that the final expressions are same as those stated by \cite{Bls9s,Bls9s_P} for bifurcated waveguides of square lattice as well as for honeycomb \cite{Bls5k_tube,Bls5c_tube,Bls5ek_tube}, in particular, 
\eqref{RTk_sq_X} can be expressed as
\begin{subequations}
\begin{eqn}
{\mathscr{R}}({{z}}_{{\text{in}}})={\mathtt{C}}_{RT}\sum\nolimits_{{{z}}\in{{\mathcal{Z}}}^+}\frac{\overline{{\mathscr{N}}_-({{z}}){\mathscr{D}}_+({{z}})}}{{\mathscr{D}}_-({{z}}){\mathscr{N}}'_+({{z}})}\frac{{{z}}_{{\text{in}}}}{({{z}}-{{z}}_{{\text{in}}})^2},
{\mathscr{T}}({{z}}_{{\text{in}}})&={\mathtt{C}}_{RT}\sum\nolimits_{{{z}}\in{{\mathcal{Z}}}^-}\frac{\overline{{\mathscr{N}}_-({{z}}){\mathscr{D}}_+({{z}})}}{{\mathscr{D}}'_-({{z}}){\mathscr{N}}_+({{z}})}\frac{{{z}}_{{\text{in}}}}{({{z}}-{{z}}_{{\text{in}}})^2},
\label{RTk_sq_X1}
\end{eqn}
\begin{eqn}
\text{where }
{\mathtt{C}}_{RT}({{z}}_{{\text{in}}})&=\frac{{{z}}_{{\text{in}}}{\mathscr{D}}_-({{z}}_{{\text{in}}}){\mathscr{N}}_+({{z}}_{{\text{in}}})}{\overline{{\mathscr{N}}'_-({{z}}_{{\text{in}}})}\overline{{\mathscr{D}}_+({{z}}_{{\text{in}}})}}.
\label{CRexp}
\end{eqn}
\label{RT_sq_com}
\end{subequations}

Analogous derivation can be carried out for the incidence when ${\mathfrak{s}}={B}$, where in place of \eqref{CRexp}, it is found that
\begin{eqn}
{\mathtt{C}}_{RT}({{z}}_{{\text{in}}})&=\frac{{{z}}_{{\text{in}}}{\mathscr{D}}_-({{z}}_{{\text{in}}}){\mathscr{N}}_+({{z}}_{{\text{in}}})}{\overline{{\mathscr{N}}_-({{z}}_{{\text{in}}})}\overline{{\mathscr{D}}'_+({{z}}_{{\text{in}}})}}.
\label{CRexp1}
\end{eqn}

\subsection{Case (e)}
Comparing with the expression of $\su_{\cdot,2}$, the coefficients ${{\mathrm{A}}}_{{{\kappa}}({{z}})}$ \eqref{modalexpansion} can be found. For example, in case of ${{\mathfrak{s}}}={A}$, neither ${z}_{{\text{in}}}$ and ${z}_{{\ast}}$ corresponding to ${\mathpzc{H}}$,
\begin{eqn}{{\mathrm{A}}}_{{{\kappa}}({{z}})}&=\frac{{{a}}_{({{{\kappa}}^{{\text{in}}}}){2}}}{{{a}}_{+({{\kappa}}({{z}})){2}}}(\frac{1}{{{z}}-{{z}}_{{\text{in}}}}-\frac{{{z}}^{-1}}{1-{z}_{\sqh}{{z}}_{{\text{in}}}}\frac{\mathcal{F}({z})}{\mathcal{F}({z}_{\sqh}^{-1})}) \frac{{\mathscr{D}}_+({{z}}){\mathscr{N}}_+({{z}}_{{\text{in}}})}{{\mathscr{N}}'_+({{z}}){\mathscr{D}}_+({{z}}_{{\text{in}}})},\\{{\mathrm{A}}}_{{{\kappa}}({{z}})}&=\frac{{{a}}_{({{{\kappa}}^{{\text{in}}}}){2}}}{{{a}}_{-({{\kappa}}({{z}}))2}}(\frac{1}{{{z}}-{{z}}_{{\text{in}}}}-\frac{{{z}}^{-1}}{1-{z}_{\sqh}{{z}}_{{\text{in}}}}\frac{\mathcal{F}({z})}{\mathcal{F}({z}_{\sqh}^{-1})}) \frac{{\mathscr{N}}_-({{z}}){\mathscr{N}}_+({{z}}_{{\text{in}}})}{{\mathscr{D}}'_-({{z}}){\mathscr{D}}_+({{z}}_{{\text{in}}})}.\end{eqn}
For ${z}={z}_{{\ast}}$, but not ${z}_{{\text{in}}}$, corresponding to ${\mathpzc{H}}=0$ (note that ${z}={z}_{{\mathpzc{h}}}^{\pm1}$ leads to ${\mathpzc{Q}}({z})=2, {\vartheta}({z})=1$ so that \eqref{Lk_sq_R} implies ${{\mathpzc{L}}}({z}_{{\mathpzc{h}}}^{\pm1})={{\mathtt{U}}_{{\mathtt{N}}-1}(1)}/({({2}-1){\mathtt{U}}_{{\mathtt{N}}-2}(1)})={{\mathtt{N}}}/({{\mathtt{N}}-1})$),
it is found that
\begin{eqn}{{\mathrm{A}}}_{{{\kappa}}({{z}})}
&=\frac{{{a}}_{({{{\kappa}}^{{\text{in}}}}){2}}}{{{a}}_{+({{\kappa}}({{z}})){2}}}(-\frac{1}{{\mathtt{N}}})\frac{({z}_{{\mathpzc{h}}}-1)}{-1+{{z}}_{{\mathpzc{h}}}^{-2}}\frac{1}{1-{z}_{\sqh}{{z}}_{{\text{in}}}}\frac{1}{\mathcal{F}({z}_{\sqh}^{-1})}{{z}}_{{\mathpzc{h}}}^{-1} \frac{{\mathscr{N}}_+({{z}}_{{\text{in}}})}{{\mathscr{D}}_+({{z}}_{{\text{in}}})},\\{{\mathrm{A}}}_{{{\kappa}}({{z}})}
&=\frac{{{a}}_{({{{\kappa}}^{{\text{in}}}}){2}}}{{{a}}_{-({{\kappa}}({{z}}))2}}\frac{1}{{\mathtt{N}}-1}\frac{({z}_{{\mathpzc{h}}}^{-1}-1)}{-1+{{z}}_{{\mathpzc{h}}}^{2}}\frac{1}{1-{z}_{\sqh}{{z}}_{{\text{in}}}}\frac{1}{\mathcal{F}({z}_{\sqh}^{-1})}{{z}}_{{\mathpzc{h}}}\frac{{\mathscr{N}}_+({{z}}_{{\text{in}}})}{{\mathscr{D}}_+({{z}}_{{\text{in}}})}.
\end{eqn}
A parallel derivation can be carried out for the incidence when ${\mathfrak{s}}={B}$.

\subsection{Case (g)}
Using \eqref{u2pmsol_gh}, by the inverse discrete Fourier transform, 
${\su}_{{\mathtt{x}}, 2}=\frac{1}{2\pi i}\oint_{\mathbb{T}} {\su}_{2;\pm}({{z}}){{z}}^{{\mathtt{x}}-1}d{{z}}, {\mathtt{x}}\in\mathbb{Z}^{\pm},$
and upon substitution
of \eqref{u1zsol_ab}, the exact expression analogous to \eqref{um1asym1} can be constructed. Further, it is found that
\begin{eqn}
{\su}_{{\mathtt{x}}, 2}
&\sim{{\mathrm{A}}}{{a}}_{({{{\kappa}}^{{\text{in}}}}){1}}\frac{{{{\mathpzc{L}}}}_+({{z}}_{{\text{in}}}){{\mathpzc{Q}}}({{z}}_{{\text{in}}})}{{{z}}_{{\text{in}}}-{{z}}_{\sq}^{-1}}
\sum\nolimits_{{{z}}\in{{\mathcal{Z}}}^+}\frac{{{z}}-{{z}}_{\sq}^{-1}}{{{z}}-{{z}}_{{\text{in}}}}\frac{{\mathscr{D}}_+({{z}})}{{\mathscr{N}}'_+({{z}})}{{z}}^{{\mathtt{x}}}, {\mathtt{x}}\to+\infty,\\
{\su}_{{\mathtt{x}}, 2}&\sim{{\mathrm{A}}}{{a}}_{({{{\kappa}}^{{\text{in}}}}){1}}\frac{{{{\mathpzc{L}}}}_+({{z}}_{{\text{in}}}){{\mathpzc{Q}}}({{z}}_{{\text{in}}})}{{{z}}_{{\text{in}}}-{{z}}_{\sq}^{-1}}
\sum\nolimits_{{{z}}\in{{\mathcal{Z}}}^-}\frac{{{z}}-{{z}}_{\sq}^{-1}}{{{z}}-{{z}}_{{\text{in}}}}\frac{{\mathscr{N}}_-({{z}})}{{\mathscr{D}}'_-({{z}})}{{z}}^{{\mathtt{x}}}\\
&-{{\mathrm{A}}}{{a}}_{({{{\kappa}}^{{\text{in}}}}){1}}(1-{{{\mathpzc{L}}}}({{z}}_{{\text{in}}})){{\mathpzc{Q}}}({{z}}_{{\text{in}}}){{z}}_{{\text{in}}}^{{\mathtt{x}}}, {\mathtt{x}}\to-\infty.
\label{wsolpm}
\end{eqn}
It is easy to see that the far-field can be determined (suitably) in terms of the eigenmodes.
It is required that \eqref{modalexpansion} holds.
Comparing the expressions of ${\su}_{{\mathtt{x}}, 2}$, 
it is found that\begin{eqn}{{\mathrm{A}}}_{{{\kappa}}({{z}})}&=\frac{{{\mathrm{A}}}{{a}}_{({{{\kappa}}^{{\text{in}}}}){1}}}{{{a}}_{+({{\kappa}}({{z}}))2}}\frac{{{{\mathpzc{L}}}}_+({{z}}_{{\text{in}}}){{\mathpzc{Q}}}({{z}}_{{\text{in}}})}{{{z}}_{{\text{in}}}-{{z}}_{\sq}^{-1}}\frac{{{z}}-{{z}}_{\sq}^{-1}}{{{z}}-{{z}}_{{\text{in}}}}\frac{{\mathscr{D}}_+({{z}})}{{\mathscr{N}}'_+({{z}})},\\{{\mathrm{A}}}_{{{\kappa}}({{z}})}&=\frac{{{\mathrm{A}}}{{a}}_{({{{\kappa}}^{{\text{in}}}}){1}}}{{{a}}_{-({{\kappa}}({{z}}))2}}\frac{{{{\mathpzc{L}}}}_+({{z}}_{{\text{in}}}){{\mathpzc{Q}}}({{z}}_{{\text{in}}})}{{{z}}_{{\text{in}}}-{{z}}_{\sq}^{-1}}\frac{{{z}}-{{z}}_{\sq}^{-1}}{{{z}}-{{z}}_{{\text{in}}}}\frac{{\mathscr{N}}_-({{z}})}{{\mathscr{D}}'_-({{z}})}.\end{eqn}
Using above equation for the coefficients ${{\mathrm{A}}}_{{{\kappa}}({{z}})}$, 
the asymptotic expression of the total displacement field follows.
Eventually \eqref{RT_sq_com} is obtained.
A derivation of the same type can be carried out for the incidence when ${\mathfrak{s}}={B}$.

\begin{figure}[ht]\centering
\includegraphics[width=.9\textwidth]{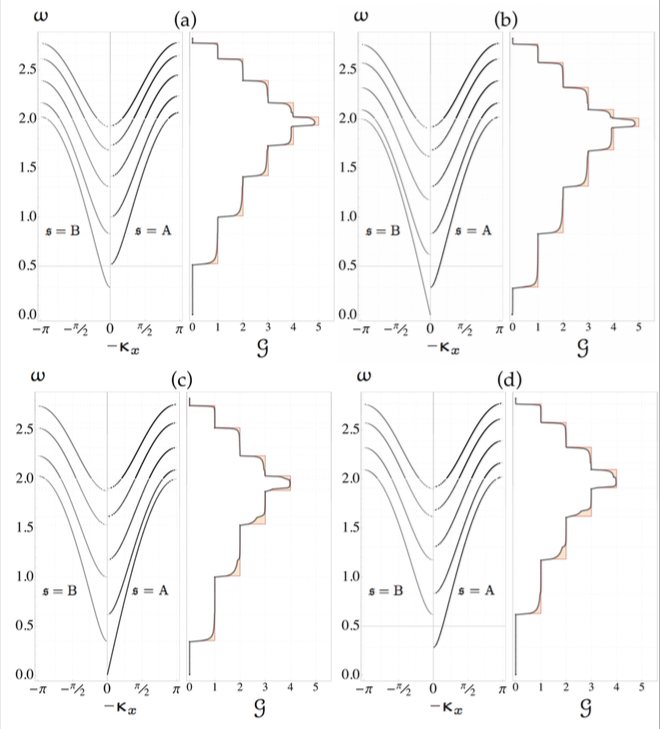}
\caption{Transmission conductance for square lattice waveguides (a)--(d).
${\mathtt{N}}=5$. 
Refer Fig. \ref{stepboundarywaveguides_sq}(a), (b), (c), (d) for the geometry of the structure.}\label{square_stepboundarywaveguides_Conduct_Cases_ad}\end{figure}

\begin{figure}[ht]\centering
\includegraphics[width=.9\textwidth]{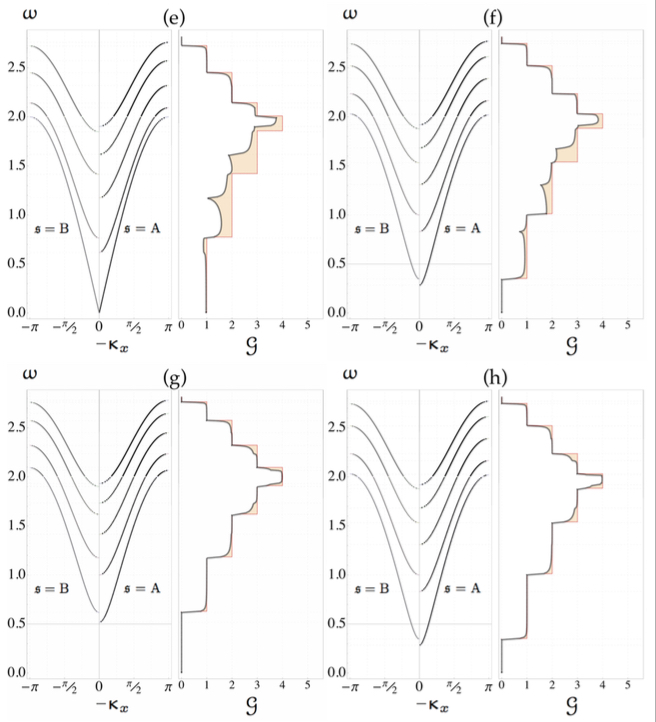}\\
\caption{Transmission conductance for square lattice waveguides (e)--(h).
${\mathtt{N}}=5$. Refer Fig. \ref{stepboundarywaveguides_sq}(e), (f), (g), (h) for the geometry of the structure.}\label{square_stepboundarywaveguides_Conduct_Cases_eh}\end{figure}

\begin{figure}[ht]\centering
\includegraphics[width=\textwidth]{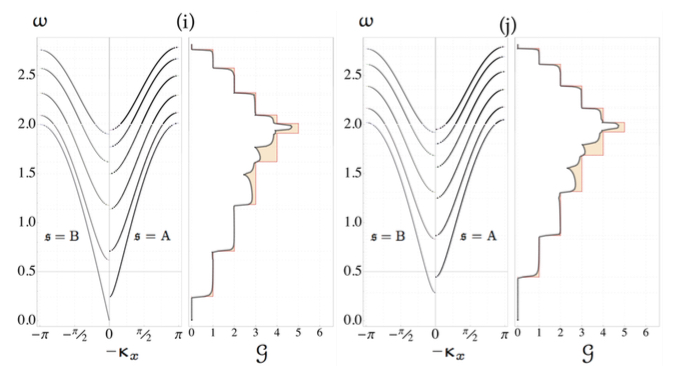}
\caption{Transmission conductance for square lattice waveguides (i) and (j).
${\mathtt{N}}=6$. Refer Fig. \ref{stepboundarywaveguides_sq}(i), (j) for the geometry of the structure.}\label{square_stepboundarywaveguides_Conduct_Cases_ij}\end{figure}

\begin{remark}
The analysis of (b) ${\mathfrak{S}\hspace{-.4ex}}{\mathbin{\substack{\circ\\\circ\bullet}}}$ follows that for ${\mathfrak{S}\hspace{-.4ex}}{\mathbin{\substack{{\bullet}\\ {\circ}{\bullet}}}}$.
The case of (c) ${\mathfrak{S}\hspace{-.4ex}}{\mathbin{\substack{\circ\\\bullet\circ}}}$ (almost) coincides with a case (denoted by ${\mathfrak{S}\hspace{-.4ex}}{\mathbin{\substack{\bullet}{\hspace{-.2ex}}\substack{\circ\\ \circ}}}$) in the square lattice bifurcated waveguides \cite{Bls9s} for ${\mathtt{N}_{\mathfrak{a}}}={\mathtt{N}}, {\mathtt{N}_{\mathfrak{b}}}=0$.
In fact, the analysis of (d) ${\mathfrak{S}\hspace{-.4ex}}{\mathbin{\substack{\bullet\\\bullet\circ}}}$ runs parallel to that for ${\mathfrak{S}\hspace{-.4ex}}{\mathbin{\substack{\circ\\\bullet\circ}}}$.
The analysis of (h) ${\mathfrak{S}\hspace{-.4ex}}{\mathbin{\substack{\circ\\\bullet\bullet}}}$ is analogous to that for ${\mathfrak{S}\hspace{-.4ex}}{\mathbin{\substack{\bullet\\\bullet\bullet}}}$.
Eventually \eqref{RT_sq_com} is obtained in all six cases (a), (b), (c), (d), (g) and (h) with the exception that ${\mathscr{D}}$ is replaced by $\mathring{{\mathscr{D}}}$ in case of the latter four cases, i.e., (c), (d), (g) and (h).
The details have been avoided in the main paper. 

The analysis of case (f), i.e., ${\mathfrak{S}\hspace{-.4ex}}{\mathbin{\substack{\bullet\\\circ\circ}}}$ is similar to that for ${\mathfrak{S}\hspace{-.4ex}}{\mathbin{\substack{\circ\\\circ\circ}}}$.
Further, it turns out that the analysis of case (i), i.e., ${\mathfrak{S}\hspace{-.4ex}}{\mathbin{\substack{\circ\\\circ\overline{\bullet}}}}$ and (j), i.e., ${\mathfrak{S}\hspace{-.4ex}}{\mathbin{\substack{{\bullet}\\ {\circ}\overline{\bullet}}}}$, of Fig. \ref{stepboundarywaveguides_sq}, closely follows that for (e) and (f).
It is crucial to note that the expression of reflectance and transmittance cannot be reduced to \eqref{RT_sq_com} in these cases.
\label{remaincases}\end{remark}

\section{Numerical results}
\label{reftransnum}
Suppose the number of propagating wave modes is $N_{{A}}$ ($=\#{{\mathcal{Z}}}^+=\#\widetilde{{\mathcal{Z}}}^-$) on the right side and $N_{{B}}$ ($=\#{{\mathcal{Z}}}^-=\#\widetilde{{\mathcal{Z}}}^+$) on the left side at a given frequency ${\upomega}$ (recall Eq. \eqref{energyflux_inc}). Then the maximum value, so called ballistic limit, is the minimum of $N_{{A}}$ and $N_{{B}}$. On the other hand, within the wave propagation equation assumed for the discrete nanoribbon structures, accounting for the evanascent and propagating wave modes excited by the presence of a step discontinuity, the transmission conductance ${\mathscr{G}}$ can be obtained. For all ten square lattice waveguides, is obtained as the sum of transmittance for all incident wave modes from one side to the other side at ${\upomega}$. In particular,
\begin{eqn}
{\mathscr{G}}={\mathscr{G}}_{{A}\leftarrow{B}}=\sum_{{{z}}_{{\text{in}}}\in\widetilde{{\mathcal{Z}}}^-}{\mathscr{T}}({{z}}_{{\text{in}}})={\mathscr{G}}_{{B}\leftarrow{A}}=\sum_{{{z}}_{{\text{in}}}\in\widetilde{{\mathcal{Z}}}^+}{\mathscr{T}}({{z}}_{{\text{in}}}).
\label{conductance_gen}\end{eqn}
The dependence of ${\mathscr{G}}$ on the frequency ${\upomega}$ is shown in 
Fig. \ref{square_stepboundarywaveguides_Conduct_Cases_ad} for cases (a)--(d),
Fig. \ref{square_stepboundarywaveguides_Conduct_Cases_eh} for cases (e)--(h),
Fig. \ref{square_stepboundarywaveguides_Conduct_Cases_ij} for cases (i) and (j), and for specific ${\mathtt{N}}$ as stated in captions. The effect of scattering is emphasized by the orange shade which demonstrates the extent to which the maximal ballistic conductance is diminished by the step.
It is quite clear that, atleast for small ${\mathtt{N}}$, that the presence of step on either free boundary or free-fixed boundary (the combination of discrete Neumann on left and discrete Dirichlet on right), see Fig. \ref{square_stepboundarywaveguides_Conduct_Cases_eh}(e), (f), and Fig. \ref{square_stepboundarywaveguides_Conduct_Cases_ij},  is associated with large effects in the middle of the pass band (for square lattice structure under study). 
As ${\mathtt{N}}$ increases, in all cases, the shaded region however reduces as the atomic sized step looses its prominence.

\begin{remark}
The fact that reflectance and transmittance sum to unity has been established as the zero lemma by \cite{Bls9s} in view of the coincidence of the expressions for cases (a), (b), (c), (d), (g), (h).
For the cases (e), (f), (i), (j), the expression of reflectance and transmittance is more involved and the proof is not applicable (few extra details for reflectance and transmittance in all cases have been also provided in the supplementary \S4); for these cases a verification of energy conservation as well as a comparison between analytical expressions and numerical solution has been carried out though the details are omitted in main paper (see supplementary \S5).
An illustration of the numerical solution of the discrete wave propagation problem based on the scheme summarized in Appendix D of \cite{Bls0} (modulo the removal of upper and lower absorbing layers, i.e. including only the left and right absorbing layers).
\label{RTcases}\end{remark}

\section{Concluding remarks}
\label{concl}
The paper presents an analysis of wave transmission across an atomic step or step-like discontinuity on any one boundary of the square lattice waveguides.
Another interpretation of the problem is that of two waveguides of infinite width and atomically different heights joined together.
The extension to wider steps (multiple semi-infinite atomic layers) brings many more complicated aspects of the mathematical analysis, besides reliance on numerics to a greater extent, however, the basic issue of effect of discreteness is anticipated to prevail in a similar manner.
It is left as an interesting mathematical exercise to see how the limit ${\mathtt{N}}\to\infty$ is approached in the context of the results of \cite{Bls10mixed}; an issue to be dealt with is the relation between incident `wave-mode' and `bulk' wave (in some way, the wave modes are linear combinations of certain bulk waves). 

From a different physical viewpoint of the electronic transport, the waveguides with discrete Dirichlet boundary condition appear naturally and some results are relevant as presented in the article (see also \cite{Bls5c_tube_media,Bls5c_tube,Bls5ek_tube}).
Using the transfer integral $\beta$ between nearest-neighbor sites, under certain assumptions, the tight-binding Hamiltonian \cite{callaway1964energy} for nanoribbons (in the second quantization) is 
${\mathcal{H}}=-\beta\sum_{\bfrac{{\mathtt{x}}}{\in\mathbb{Z}}}(\sum_{{{\mathtt{y}}}{\in\mathbb{Z}_1^{{\mathtt{N}}-1}}}(a_{{\mathtt{x}}}^\dagger({\mathtt{y}})a_{{{\mathtt{x}}}-1}({\mathtt{y}})+a_{{\mathtt{x}}}^\dagger({\mathtt{y}}+1)a_{{{\mathtt{x}}}}({\mathtt{y}})))+h.c., $
where $a^\dagger_{{\mathtt{x}}}({\mathtt{y}})$ is the {creation} operator of an electron at the $({\mathtt{x}}, {\mathtt{y}})$ site; $a_{{\mathtt{x}}}({\mathtt{y}})$ is the corresponding {annihilation} operator. 
After an application of the discrete Fourier transform
and expanding the single-particle electronic wave functions, the Schr{\"{o}}dinger equation ${\mathcal{H}}({{z}})|\Psi({{z}})\rangle= {{\mathcal{E}}}|\Psi({{z}})\rangle,$ leads to an association with the mechanical model via
$-\beta^{-1}{{\mathcal{E}}}{\su}_{{{\mathtt{y}}}}=({z}+{z}^{-1}){\su}_{{{\mathtt{y}}}}+{\su}_{{{{\mathtt{y}}}}-1}+{\su}_{{{{\mathtt{y}}}}+1}.$ 

\section*{Acknowledgement}
The partial support of SERB MATRICS grant MTR/2017/000013 is gratefully acknowledged.

\renewcommand*{\bibfont}{\footnotesize}
\printbibliography

\begin{appendix}
\section{Auxiliary expressions}
\label{app_recallbifwavesq}

\subsection{}
\label{app_recallbifwavesq_Chebdef}
It is easily conceivable that there are four combinatorial variants of the square lattice waveguides (also the same are studied in detail by \cite{Bls9}) which are denoted by self-explanatory notation ${\mathfrak{S}{\sbfrac{\bullet}{\bullet}}},{\mathfrak{S}{\sbfrac{\bullet}{\circ}}},{\mathfrak{S}{\sbfrac{\circ}{\bullet}}},{\mathfrak{S}{\sbfrac{\circ}{\circ}}}$ where the superscript on ${\mathfrak{S}}$ represents the upper boundary and the subscript represents the lower boundary.
Indeed, in terms of the Chebyshev polynomials, using the definition \eqref{chebx_sq}, the dispersion relations for square lattice waveguides of width ${\mathtt{N}}$ \cite{Bls9,Bls9s} are:
\begin{subequations}
\begingroup
\addtolength{\jot}{0em}
\begin{align}
\text{ for }{\mathfrak{S}{\sbfrac{\bullet}{\bullet}}}&&{F}^{{\mathtt{N}}}_{{\mathfrak{S}{\sbfrac{\bullet}{\bullet}}}}={\mathtt{U}}_{{\mathtt{N}}}({\vartheta})&=0,\label{disp_sq_X}\\
\text{ for }{\mathfrak{S}{\sbfrac{\circ}{\circ}}}&&{F}^{{\mathtt{N}}}_{{\mathfrak{S}{\sbfrac{\circ}{\circ}}}}={\mathpzc{H}}{\mathtt{U}}_{{\mathtt{N}}-1}({\vartheta})&=0,
\label{disp_sq_R}\\
\text{ for }{\mathfrak{S}{\sbfrac{\bullet}{\circ}}}, {\mathfrak{S}{\sbfrac{\circ}{\bullet}}}&&{F}^{{\mathtt{N}}}_{{\mathfrak{S}{\sbfrac{\bullet}{\circ}}}}={F}^{{\mathtt{N}}}_{{\mathfrak{S}{\sbfrac{\circ}{\bullet}}}}={\mathtt{U}}_{{\mathtt{N}}}({\vartheta})-{\mathtt{U}}_{{\mathtt{N}}-1}({\vartheta})&=0.
\label{disp_sq_XR}
\end{align}
\endgroup
\end{subequations}
Also \eqref{disp_sq_XR}, using the Chebyshev polynomials ${\mathtt{V}}_n$ of the third kind, can be expressed as ${\mathtt{V}}_{{\mathtt{N}}}({\vartheta})=0$. Moreover, 
the expressions of the numerator and denominator of the kernel can be further expanded as stated in Table \ref{stripbc_sq_ND}; in particular, 
\eqref{Lk_sq_X} transforms to
${{\mathpzc{L}}}={{F}^{{\mathtt{N}}}_{{\mathfrak{S}{\sbfrac{\bullet}{\bullet}}}}}/{{F}^{{\mathtt{N}}}_{{\mathfrak{S}{\sbfrac{\bullet}{\circ}}}}},$
\eqref{Lk_sq_RX} to
${{\mathpzc{L}}}={{F}^{{\mathtt{N}}}_{{\mathfrak{S}{\sbfrac{\circ}{\bullet}}}}}/{{F}^{{\mathtt{N}}}_{{\mathfrak{S}{\sbfrac{\circ}{\circ}}}}},$
\eqref{Lc_sq_R} to
${{\mathpzc{L}}}
={{F}^{{\mathtt{N}}}_{{\mathfrak{S}{\sbfrac{\circ}{\circ}}}}}/{({\mathpzc{Q}}-1){F}^{{\mathtt{N}}-1}_{{\mathfrak{S}{\sbfrac{\circ}{\bullet}}}}},$
\eqref{Lc_sq_XR} to
${{\mathpzc{L}}}
={{F}^{{\mathtt{N}}}_{{\mathfrak{S}{\sbfrac{\bullet}{\circ}}}}}/{({\mathpzc{Q}}-1){F}^{{\mathtt{N}}-1}_{{\mathfrak{S}{\sbfrac{\bullet}{\bullet}}}}},$
\eqref{Lk_sq_R} to
${{\mathpzc{L}}}
={{F}^{{\mathtt{N}}}_{{\mathfrak{S}{\sbfrac{\circ}{\circ}}}}}/{({\mathpzc{Q}}-1){F}^{{\mathtt{N}}-1}_{{\mathfrak{S}{\sbfrac{\circ}{\circ}}}}},$
\eqref{Lk_sq_XR} to
${{\mathpzc{L}}}
={{F}^{{\mathtt{N}}}_{{\mathfrak{S}{\sbfrac{\bullet}{\circ}}}}}/{({\mathpzc{Q}}-1){F}^{{\mathtt{N}}-1}_{{\mathfrak{S}{\sbfrac{\bullet}{\circ}}}}},$
\eqref{Lc_sq_X} to
${{\mathpzc{L}}}
={{F}^{{\mathtt{N}}}_{{\mathfrak{S}{\sbfrac{\bullet}{\bullet}}}}}/{{\mathpzc{Q}}{F}^{{\mathtt{N}}-1}_{{\mathfrak{S}{\sbfrac{\bullet}{\bullet}}}}},$
\eqref{Lc_sq_RX} to
${{\mathpzc{L}}}
={{F}^{{\mathtt{N}}}_{{\mathfrak{S}{\sbfrac{\circ}{\bullet}}}}}/{{\mathpzc{Q}}{F}^{{\mathtt{N}}-1}_{{\mathfrak{S}{\sbfrac{\circ}{\bullet}}}}},$
\eqref{Lk_sq_RX2} to
${{\mathpzc{L}}}
={{F}^{{\mathtt{N}}}_{{\mathfrak{S}{\sbfrac{\circ}{\bullet}}}}}/{{\mathpzc{Q}}{F}^{{\mathtt{N}}-1}_{{\mathfrak{S}{\sbfrac{\circ}{\circ}}}}},$
and \eqref{Lk_sq_X2} to
${{\mathpzc{L}}}
={{F}^{{\mathtt{N}}}_{{\mathfrak{S}{\sbfrac{\bullet}{\bullet}}}}}/{{\mathpzc{Q}}{F}^{{\mathtt{N}}-1}_{{\mathfrak{S}{\sbfrac{\bullet}{\circ}}}}}.$

\subsection{}
By application of the discrete Fourier transform \eqref{unpm}, the discrete Helmholtz equation \eqref{dHelmholtz_sq}, for all ${\mathtt{y}}\in\mathbb{Z}$ with ${\mathtt{y}}$ away from the boundary of the given lattice waveguide, is expressed 
as
\beqans
{{\mathpzc{Q}}}({{z}}){\su}_{{\mathtt{y}}}^F({{z}})&-&({\su}_{{\mathtt{y}}+1}^F({{z}})+{\su}_{{\mathtt{y}}-1}^F({{z}}))=0, \label{dHelmholtzF_sq}\\
\text{where }{{\mathpzc{Q}}}({{z}})&{:=}&4-{{z}}-{{z}}^{-1}-{\upomega}^2, \label{q2_sq}\\
{{\lambda}}&{:=}&\frac{{{\mathpzc{r}}}-{{\mathpzc{h}}}}{{{\mathpzc{r}}}+{{\mathpzc{h}}}}, {{\mathpzc{h}}}{:=}\sqrt{{\mathpzc{H}}}, {{\mathpzc{r}}}{:=}\sqrt{{\mathpzc{R}}},\label{lambdadef_sq}\\
{{\mathpzc{H}}}&{:=}& {\mathpzc{Q}}-2, {{\mathpzc{R}}}{:=} {\mathpzc{Q}}+2.\label{HR_sq}
\eeqans{dHelmholtzF_sq_full}
The complex functions ${\mathpzc{h}}, {\mathpzc{r}}, {{\mathpzc{H}}}$, ${{\mathpzc{R}}}$, and ${\lambda}$ are defined on $\mathbb{C}\setminus{\mathscr{B}}$ where ${\mathscr{B}}$ denotes the union of branch cuts for ${{\lambda}}$, borne out of the chosen branch
$-\pi<\arg {{\mathpzc{H}}}({{z}}) <\pi, \Re {{\mathpzc{h}}}({{z}})>0, \Re {{\mathpzc{r}}}({{z}})>0, {\text{\rm sgn}} \Im {{\mathpzc{h}}}({{z}})={\text{\rm sgn}} \Im {{\mathpzc{r}}}({{z}}),$
for ${{\mathpzc{h}}}$ and ${{\mathpzc{r}}}$ such that 
$|{{\lambda}}({{z}})|\le1, {{z}}\in\mathbb{C}\setminus{\mathscr{B}},$ as ${\upomega}_2$ in \eqref{lambdadef_sq} is positive. Note that ${\lambda}+{\lambda}^{-1}={\mathpzc{Q}}$.
The general solution of \eqref{dHelmholtzF_sq} is given by the expression
\begin{eqn}
{\su}_{{\mathtt{y}}}^F={\mathit{c}}_1 {\lambda}^{{\mathtt{y}}}+{\mathit{c}}_2 {\lambda}^{-{\mathtt{y}}},
\label{gensol_sq}
\end{eqn}
where ${\mathit{c}}_{1, 2}$ are arbitrary analytic functions of ${z}$ in ${{\mathscr{A}}}$ (to be specified in the Wiener--Hopf formulation for each case).
\end{appendix}
\end{document}